\documentclass[a4paper,11pt]{article}
\usepackage{jheppub} % for details on the use of the package, please see the JINST-author-manual
\arxivnumber{2507.00494} % if you have one

\usepackage{graphicx,bm,color,wrapfig,cancel}
\usepackage{amsmath,amssymb,bm}
\usepackage{float}
\usepackage{braket}
\usepackage{comment}
\usepackage{subfigure}
\usepackage{slashed}

\usepackage{tikz}
\usetikzlibrary{matrix}

\newcommand{\fulltoday}{\number\day\space \ifcase\month\or
    January\or February\or March\or April\or May\or June\or
    July\or August\or September\or October\or November\or December\fi
    \space\number\year}
 \makeatletter
    
    \@addtoreset{equation}{section}
  \makeatother

\allowdisplaybreaks[3]

\title{\boldmath
Susceptibilities of rotating quark matter in Fourier-Bessel basis
}

\author[a]{Mamiya~Kawaguchi,}
\author[b,c]{Kazuya~Mameda}

\affiliation[a]{Center for Fundamental Physics, School of Mechanics and Physics, Anhui University of Science and Technology, Huainan, Anhui 232001, People’s Republic of China}

\affiliation[b]{
Department of Physics, Tokyo University of Science, Tokyo 162-8601, Japan}

\affiliation[c]{
RIKEN iTHEMS, RIKEN, Wako 351-0198, Japan }

\emailAdd{mamiya@aust.edu.cn}
\abstract{
We analyze various two-point correlation functions of fermionic bilinears in a rotating finite-size cylinder at finite temperatures, with a focus on susceptibility functions. Due to the noninvariance of radial translation, the susceptibility functions are constructed using the Dirac propagator in the Fourier-Bessel basis instead of the plane-wave basis. As a specific model to demonstrate the susceptibility functions in an interacting theory, we employ the two-flavor Nambu–Jona-Lasinio model. We show that the incompatibility between the mean-field analysis and the Fourier-Bessel basis is evaded under the local density approximation, and derive the resummation formulas of susceptibilities with the help of a Ward-Takahashi identity. The resulting formulation reveals the rotational effects on meson, baryon number, and topological susceptibilities, as well as the moment of inertia. Our results may serve a useful benchmark for future lattice QCD simulations in rotating frames.
}

\begin{document}

\maketitle
\flushbottom

%\preprint{RIKEN-iTHEMS-Report-25}

%%%%%%%%%%%%%%%%%%%%%%%%%%%%%%%%%%%%%%%%%%%%
\section{Introduction}

Rotating QCD matter has garnered significant attention in recent years, particularly  in the context of heavy-ion collisions, which create the most vortical fluid known in nature~\cite{STAR:2017ckg}.
On the theoretical side, the rigidly rotating system is an idealized setup in order to understand fundamental features of rotating relativistic matter~\cite{Vilenkin:1978hb,Vilenkin:1979ui,Vilenkin:1980zv}, and has already revealed typical rotational responses of QCD matter in terms of the chiral symmetry breaking~\cite{Chen:2015hfc,Jiang:2016wvv, Ebihara:2016fwa,Chernodub:2016kxh,Wang:2018sur,Chen:2023cjt,Gaspar:2023nqk,Nunes:2024hzy}, the deconfinement transition~\cite{Chernodub:2020qah,Chen:2020ath,Fujimoto:2021xix,Braga:2022yfe,Zhao:2022uxc,Yadav:2022qcl,Chen:2022smf,Chen:2024tkr,Mameda:2023sst,Jiang:2023zzu,Chen:2024edy}, both of them~\cite{Sun:2023kuu,Sun:2024anu,Chen:2024jet,Cao:2023olg}, and other possible transitions~\cite{Huang:2017pqe,Liu:2017spl,Cao:2019ctl,Chen:2019tcp,Zhang:2018ome};
see also the corresponding lattice studies in Ref.~\cite{Chernodub:2022veq, Braguta:2021jgn, Braguta:2022str, Braguta:2023yjn,Braguta:2023iyx,Yang:2023vsw,Wang:2025mmv,Fukushima:2025hmh}.
Even under such simplification, however, accessible aspects of rotating QCD matter are still limited.
% Although there are less effective-model approaches of rotating matter compatible with loop-diagrams,
In particular, since the first principle simulation under rotation suffers from a sign problem~\cite{Yamamoto:2013zwa}, the sophistication of the model approaches and perturbative computation is an inevitable direction.

A difficulty we face in field-theoretical studies of relativistic rotating matter is the coordinate-dependence of physical observables (see also Refs.~\cite{Chen:2015hfc,Fukushima:2024tkz} for an exceptional case under a strong magnetic field).
As demonstrated in preceding works, maintaining causality in a relativistic rotating system necessitates the consideration of a finite-size geometry, typically a cylinder~\cite{Ambrus:2014uqa}.
Such a finite-size system violates translational invariance along the radial direction, leading to coordinate-dependent thermodynamic quantities.
In more field-theoretical language, propagators of rotating matter are described not by the plane-wave basis, but the Fourier-Bessel basis.
As a result, in an interacting theory under rotation, it is necessary to carefully revisit to standard schemes employed in the usual nonrotating case, including the Feynman rules~\cite{Kuboniwa:2025vpg}. 

A typical modification is found in effective models describing phase transition under rotation.
The standard mean-field analysis is not directly applied to radially inhomogeneous order parameters.
One way to bypass this issue is to utilize the local density approximation, where the radial modulation of the order parameter is supposed to be negligible.
Under this approximation, for instance, the mean-field analysis for the spontaneous chiral symmetry breaking can be performed parallelly to the usual case, but the constant dynamical mass is replaced with the radial-coordinate dependent one~\cite{Jiang:2016wvv}.
A remarkable feature of this analysis is that the self-consistent gap equation involves only the single radial coordinate (through the one-loop of a single propagator), and this is the reason why such a simple analysis works.
However, the validity of this strategy becomes subtle when we argue mesonic fluctuations and susceptibilities, which consist of multi-point propagators, and thus involve more radial coordinates than one.

In order to deal with such an issue, Ward-Takahashi (WT) identities would provide a key computational guide.
WT identities are highly useful in investigations of QCD properties at finite temperature and density, where Lorentz symmetry is violated by external conditions.
For instance, a WT identity in term of the chiral symmetry relates the quark condensate to the two-point correlation function derived from quark bilinear operators corresponding to the pion mode (referred to as the pion susceptibility), and thus plays a crucial role in assessing the consistency within an effective model approach~\cite{GomezNicola:2019myi,Cui:2021bqf,Cui:2022vsr,Kawaguchi:2023olk}.
Furthermore, a WT identity for the U(1) axial symmetry associates quark two-point correlation functions with the topological susceptibility, which is originally described by gluonic degrees of freedom~\cite{GomezNicola:2016ssy,GomezNicola:2017bhm,GomezNicola:2019myi,Kawaguchi:2020qvg,Cui:2021bqf,Cui:2022vsr}.
Since these WT identities related to the global symmetries is irrelevant to the spacetime geometry, they are expected to be powerful constraints even in the quantum field theory for rotating matter.

The present work aims to analyze fermionic two-point correlation functions of rotating matter in the Fourier-Bessel basis, within the mean-field level of an Nambu-Jona-Lasinio (NJL) model.
We compute various quark susceptibilities, including meson, baryon, and topological susceptibilities.
In a similar computational manner, we can evaluate the {\it density} of moment of inertia in the interacting theory, while the net moment of inertia has already derived in several preceding studies~\cite{Chernodub:2016kxh,Sun:2023yux,Ahadi:2025rqs}.
Besides, we show that a WT identity enforces the mesonic susceptibilities to take the similar form to that in the translational invariant systems.
These analyses not only establish the groundwork for more comprehensive treatments of quantum fluctuations in rotating matter, but also provides new benchmarks to characterize the rotational response to QCD phase transition.
Indeed, without rotation, the phase structure of hot QCD has been explored, revealing the chiral crossover where the quark condensate decreases smoothly with increasing temperature~\cite{Aoki:2006we,Bhattacharya:2014ara}.
Additionally, meson susceptibilities, derived from quark bilinear correlation functions, have also been investigated and have provided valuable insights into meson property fluctuations and symmetry breaking, including chiral $SU(2)$ and $U(1)$ axial symmetries~\cite{HotQCD:2012vvd,Bhattacharya:2014ara,Buchoff:2013nra,Aoki:2020noz,Aoki:2021qws}.
Given this fact, our study extends the understanding of rotating QCD matter by elucidating how rotation modifies these susceptibility functions and the associated symmetry restoration patterns.

This paper is organized as follows.
In section~\ref{setup}, we present the theoretical framework in the rotating cylindrical frame.
We briefly review the formulation of the Dirac propagator described in the Fourier-Bessel basis and introduce the definition of the susceptibility functions.
The boundary condition in a finite cylinder is also provided. 
In section~\ref{sus_NJL_model}, we analyze the susceptibility functions within the NJL model under the mean-field approximation. 
We examine the implementation of the local density approximation in the current framework and explore the associated constraints on the quark propagator. This approximation simplifies the calculation of the susceptibility functions and, guided by a WT identity, makes it possible to evaluate meson susceptibilities.
In section~\ref{numerical_sus}, we show our numerical results on the susceptibility functions at finite temperature under rotation.
Finally, we summarize our findings and discuss possible future directions in section~\ref{summary}.

%%%%%%%%%%%%%%%%%%%%%%%%%%%%%%%%%%%%%%%%%%%%

\section{Theoretical setup}
\label{setup}

In this study, we aim to develop a systematic framework for calculating fermionic two-point correlation functions of rotating matter in the Fourier-Bessel basis. 
To facilitate this development, we take several preparatory steps in this section.
Specifically, we first introduce the Dirac propagator in a rotating frame. Furthermore, we provide a clear definition of susceptibility functions, which are defined by the two-point correlation function of the fermionic bilinear operators. 
To ensure that relativistic causality is preserved,
we treat the rotating frame as a finite cylindrical system and specify the corresponding boundary conditions.

%%%%%%%%%%%%%%%%%%%%%%%%%%%%%%%%%%%%%%%%%%%%
\subsection{Dirac propagator in the Fourier-Bessel basis}

In this subsection, we review the analytical expression for the Dirac propagator in a rotating frame, which is formulated by using the Fourier-Bessel basis. 
In curved spacetime, the Dirac equation is expressed as follows: 
\begin{equation}
\Bigl[i\gamma^\mu (\partial_\mu +\Gamma_\mu) -m\Bigr]\psi = 0,
\end{equation}
where $\Gamma_\mu = -\frac{i}{4}\omega_{\mu ij}\sigma^{ij} $
denotes 
the affine connection
with the Dirac spin matrices $\sigma^{ij } = \frac{i}{2}[\gamma^i, \gamma^j]$ and the spin connection 
$\omega_{\mu ij} = g_{\alpha \beta } e_i^\alpha
(\partial_\mu e_j^\beta +\Gamma^\beta_{\mu\nu} e_j^\nu  )$ with
the Christoffel symbol
$\Gamma^\beta_{\mu\nu}$ and a vierbein $e_i^\mu$.
The Greek (Latin) indices, $\mu=t,x,y,z$ ($i=0,1,2,3$), represent the coordinates in the rotating (local Lorentz) frame.
By using the vierbein $e_i^\mu$, the metric in the rotating frame is related to the Minkowski metric $\eta_{ij}$ as
\begin{equation}
g_{\mu\nu} = 
\eta_{ij} e^i_\mu
e^j_\nu.
\end{equation}

In this paper, we consider a system rotating around the $z$-axis with an angular velocity, ${\bm \Omega}= \Omega \hat {\bm z}$, and thus employ the corresponding metric as
\begin{eqnarray}
g_{\mu\nu} =
\begin{pmatrix} 
1- (x^2 + y^2)\Omega^2 & y\Omega & - x \Omega & 0\\
y\Omega &-1 &0 & 0 \\
-x\Omega &0 & -1 & 0 \\
0& 0 & 0 & -1 
\end{pmatrix}.
\end{eqnarray}
Also we choose the vierbein to be 
\begin{equation}
    e^{t}_0 = e^{x}_1 = e^y_2 = e^z_3 =1,
    \quad
    e^x_0= y\Omega,
    \quad
    e^y_0 = -x \Omega, 
\end{equation}
and the other components are zero.
Then, the Dirac equation in the rotating frame can be written as
\begin{equation}
\begin{split}
\label{eq:Dirac_eq_rot}
(i\slashed{D}-m)\psi = 0 ,\qquad
i\slashed{D} 
= \gamma^0\left\{i\partial_t + \Omega \left(-i\partial_\theta+
\frac{1}{2}
 \sigma^{12}\right)  \right\} 
+\gamma^r i \partial_r 
+\gamma^\theta \frac{i}{r}\partial_\theta
+\gamma^3 i \partial_z,
\end{split}
\end{equation}
where we use the cylindrical coordinate $r=\sqrt{x^2+y^2}$ and $\theta = \arctan (y/x)$ with the gamma matrices $\gamma^r = \gamma^1\cos\theta + \gamma^2 \sin\theta$ and $\gamma^\theta = -\gamma^1 \sin\theta + \gamma^2 \cos\theta$. 
The unnormalized linearly independent solution of this Dirac equation is given as the product of the plane-wave $e^{-ip^t t+ip_z z}$ and either of the following two functions:
\begin{equation}
\begin{split}
\Phi_{l,+}(p_\perp; \boldsymbol{x}_\perp) &= e^{il\theta}  J_l(p_\perp r)
{\cal P}_+,\\
\Phi_{l,-}(p_\perp; \boldsymbol{x}_\perp) &= e^{i(l+1)\theta   } J_{l+1}(p_\perp r)
{\cal P}_-,
\end{split}
\end{equation}
where $\boldsymbol{x}_\perp = (r,\theta)$ is the transverse coordinate, $J_l(x)$ is the $l$-th Bessel function of the first kind with the radial momentum $p_\perp$, and $\mathcal{P}_\pm = (1\pm i\gamma^1 \gamma^2)/2$ is the spin-projection operator.
One can check that $(-i\partial_\theta + \sigma^{12}/2) \Phi_{l,\pm} = (l+1/2)\Phi_{l,\pm}$, where $j=l+1/2$ corresponds to the total momentum.
This eigenvalue equation is due to the rotational symmetry around the $z$ axis.
Besides, the energy dispersion relation is given by
$p^t=\pm E_p -\Omega\left(l+1/2\right)$
with $E_p= \sqrt{p_z^2+p_\perp^2+m^2}$.

The Fourier transformation applied to the Dirac field is modified by the Bessel function as follows:
\begin{equation}
\begin{split}
\psi (x) &=
\int \frac{d^4p}{(2\pi)^4}
e^{-ip^t t +ip_z z} \sum_{l=-\infty}^\infty 
\Bigl[ 
\Phi_{l,+}
(p_\perp; \boldsymbol{x}_\perp)
+ 
\Phi_{l,-}
(p_\perp; \boldsymbol{x}_\perp)
\Bigr]\psi_l(p)
,\\
\psi_l(p)&= \int d^4x 
e^{ip^t t -ip_z z}
\Bigl[ \Phi_{l,+}^\dagger(p_\perp; \boldsymbol{x}_\perp) + 
\Phi_{l,-}^\dagger(p_\perp; \boldsymbol{x}_\perp)
\Bigr] \psi(x).
\label{FT_mod}
\end{split}
\end{equation}
The modified transformation in Eq.~\eqref{FT_mod} is referred to as the Fourier-Bessel transformation.
In the Fourier-Bessel basis, the four-dimensional delta function is expressed as  
\begin{equation}
\begin{split}
\delta^{(4)}(x-x^\prime)
&=
\int \frac{d^4 p}{(2\pi)^4} 
e^{-ip^t(t-t^\prime) +i p_z (z-z^\prime)} \\
&\quad\times
\sum_{l=-\infty}^\infty 
\left[
\Phi_{l,+}
(p_\perp; \boldsymbol{x}_\perp)
\Phi_{l,+}^\dagger
(p_\perp; \boldsymbol{x}_\perp')
+
\Phi_{l,-}
(p_\perp; \boldsymbol{x}_\perp)
\Phi_{l,-}^\dagger
(p_\perp; \boldsymbol{x}_\perp')
\right].
\label{delta_mod}
\end{split}
\end{equation}
The Dirac
propagator in the rotating frame, 
$S(x,x') = \langle0|
T\psi(x)\bar \psi(x^\prime) |0 \rangle$,
satisfies the following equation,
\begin{equation}
(i\slashed{D}_x-m)
S(x,x') 
=
i \delta^{(4)}(x-x^\prime).
\label{satisfy}
\end{equation}
This is solved by
\begin{equation}
S(x,x')
=
i \sum_{l=-\infty}^\infty 
\int \frac{d^4 p}{(2\pi)^4} 
e^{-ip^t(t-t^{\prime }) +i p_z (z-z^{\prime})}
\mathcal{S}_l(p;\boldsymbol{x}_\perp,\boldsymbol{x}_\perp'),
\label{propagator}
\end{equation}
with
\begin{equation}
\begin{split}
\mathcal{S}_l(p;\boldsymbol{x}_\perp,\boldsymbol{x}_\perp')
&=
\mathcal{S}_{l,\parallel}(p)
\Bigl[
\Phi_{l,+}^\dagger 
(p_\perp; \boldsymbol{x}_\perp')
\Phi_{l,+}
(p_\perp; \boldsymbol{x}_\perp)
+
\Phi_{l,-}^\dagger
(p_\perp; \boldsymbol{x}_\perp')
\Phi_{l,-} (p_\perp; \boldsymbol{x}_\perp)  
\Bigr]
\\
&\quad
+
\mathcal{S}_{l,\perp}(p)
\Bigl[
\Phi_{l,+}^\dagger
(p_\perp; \boldsymbol{x}_\perp')
\Phi_{l,-} 
(p_\perp; \boldsymbol{x}_\perp)
-\Phi_{l,-}^\dagger
(p_\perp; \boldsymbol{x}_\perp')
\Phi_{l,+} 
(p_\perp;\boldsymbol{x}_\perp)
\Bigr],
\end{split}
\end{equation}
\begin{equation}
 \mathcal{S}_{\parallel}(p) =
\frac{\left(p^t + \Omega l+\frac{1}{2} \Omega\right)\gamma^0
-p_z  \gamma^3
+m }{\left\{
p^t+\Omega\left(l + \frac{1}{2}\right)
\right\}^2-E_p^2
+ i\epsilon
}
    ,\qquad
\mathcal{S}_{\perp}(p) =
\frac{
-i p_\perp \gamma^1
}{\left\{
p^t+\Omega\left(l + \frac{1}{2}\right)
\right\}^2-E_p^2
+ i\epsilon
}.
\end{equation}
The Dirac propagator obtained above is derived in Ref.~\cite{Ambrus:2015lfr}.
This propagator described with the Fourier-Bessel basis reflects the violation of translational symmetry in various quantities in terms of rotating Dirac fields.

%%%%%%%%%%%%%%%%%%%%%%%%%%
\subsection{Definition of susceptibility functions}
\label{sus_def}
Before proceeding with the explicit calculation of correlation functions in the rotating frame using the Dirac propagator in Eq.~\eqref{propagator}, we provide a brief review in this subsection of how susceptibility functions are defined in field theory. 
To ensure that the susceptibility functions can be defined in any interacting system, we 
start with the general case of the Lagrangian including an appropriate interaction ${\cal L}_{\rm int}$, as follows:
\begin{equation}
\begin{split}
&\mathcal{L} = \mathcal{L}_0 -\bar \psi \zeta_{\rm sp}\psi + \mathcal{L}_\mathrm{int}, 
\qquad
\mathcal{L}_0 = \bar \psi(i\slashed{D}
 - m)\psi.
\end{split}
\end{equation}
Here, for instance, in order to define the susceptibility functions associated with the scalar and pseudoscalar fermionic-bilinear operators of the Dirac field in a systematic way, we have introduced a spurion field $\zeta_{\rm sp}$, which takes the form of
\begin{eqnarray}
\zeta_{\rm sp} = {\zeta_S}+i{\zeta_P}\gamma_5, 
\end{eqnarray}
with ${\zeta_S} ({\zeta_P})$ being the scalar (pseudoscalar) source field. 
These source fields are eventually set to zero
when evaluating physical observables.
The generating functional of this action,
including the source fields,
in the path-integral formulation is given by
\begin{eqnarray}
Z&=&
\int \mathcal{D}\bar \psi \mathcal{D}\psi
\exp\left[ i\int d^4x \,\mathcal{L}
\right]
,
\end{eqnarray}
and the corresponding effective action is evaluated as
\begin{eqnarray}
\Gamma = -i\ln Z.
\end{eqnarray}
By taking the derivative of this effective action with respect to the source fields $\zeta_S$ and $\zeta_P$,
one can obtain the susceptibility functions, 
\begin{equation}
\label{chi_def}
\int d^4x\,
\chi_{c}(x)
=
(-i)
\frac{\partial^2\Gamma}{\partial {\zeta_{c} }^2 
 }
\Biggl{|}_{\zeta_{\rm sp}=0},
\qquad 
c = S, P,
\end{equation}
where the susceptibility functions are expressed as the two-point correlation function of the scalar and pseudoscalar operators,
\begin{equation}
\label{SP_sus}
\chi_{c}(x) =
\int d^4x^\prime
\langle 0| T 
{\cal O}_{c}(x)
{\cal O}_{c}(x^\prime)
|0\rangle,
\end{equation}
with the scalar and pseudoscalar operators constructed from the fermionic-bilinear fields as
\begin{equation}
\begin{split}
{\cal O}_{S} =
\bar \psi \psi,\quad
{\cal O}_{P}
=\bar \psi 
i\gamma_5
\psi.
\end{split}
\end{equation}
The susceptibility functions in Eq.~\eqref{SP_sus} represent the quantum fluctuations of the scalar and pseudoscalar fermionic-bilinear fields at zero external momentum.

When deriving the susceptibility in Eq.~\eqref{chi_def} from the effective action, it always involves a time-space integral. 
In the usual translationally invariant case,
the susceptibility function~\eqref{SP_sus} is constant and 
thus 
Eq.~\eqref{chi_def} is reduced to Eq.~\eqref{SP_sus} multiplied by the four-dimensional volume factor.
However, it should be noted that when translational invariance is violated, this reduction is not allowed, requiring a careful treatment of the four-dimensional integral.

%%%%%%%%%%%%%%%%%%%%%%%%%%
\subsection{Boundary conditions in finite cylinder}
We argue the necessity of the boundary condition in rotating systems.
In the rotating frame with a finite angular velocity $\Omega$, a particle at the radial coordinate $r$ has the velocity $v=\Omega r$. 
To avoid an unphysical rotating system where $v$ exceeds the speed of light, an appropriate boundary condition should be imposed. 
Specifically, in the finite-size cylinder with a finite radius $R$, the maximum value on the edge $v_{\rm edge}=\Omega R$ should satisfy the causality constraint: $v_{\rm edge}\leq 1$.
In this study, we impose the boundary condition of the no-incoming flux for the fermions, ensuring that the vector current is conserved within the finite-size cylinder,
\begin{eqnarray}
R\int_{-\infty}^\infty dz
\int_0^{2\pi} d\theta \,
\bar \psi 
\gamma^r
\psi\Bigl|_{r=R} =0.
\label{no_incoming}
\end{eqnarray}
A sufficient condition for Eq.~\eqref{no_incoming} is that the radial momentum $p_\perp$ is discretized, as follows~\cite{Hortacsu:1980kv,Ambrus:2015lfr}:
\begin{equation}
\label{eq:pperp}
p_\perp
\to
p_{l,k} =
\begin{cases}
\xi_{l,k}/R \quad
&{\rm for} \quad l=0,1,\cdots,\\
\xi_{-l-1,k}/R \quad &{\rm for}\quad l=-1,-2,\cdots,
\end{cases}
\end{equation}
where $\xi_{l,k}$ represents the $k$th positive zero of $J_l(x)$.
This discretized momentum modifies the three-dimensional momentum phase space as
\begin{equation}
\sum_{l=-\infty}^\infty
\int \frac{d^3 p}{(2\pi)^3} 
%F^{(l)}(p)
\to 
\int\frac{dp_z}{2\pi}
\sum_{k=1}^\infty 
 \sum_{l=-\infty}^\infty
 \frac{1}{\pi R^2N_{l,k}^2},
 \qquad
 N_{l,k}
 =
 \begin{cases}
  |J_{l+1}(\xi_{l,k})|, & l=0,1,\cdots, \\
  |J_{-l}(\xi_{-l-1,k})|, & l=-1,-2,\cdots .
 \end{cases}
\label{phase_space}
\end{equation}
We note that the discretized momentum~\eqref{eq:pperp} and the resulting normalization factor $N_{l,k}$ are invariant under the replacement $j=l+1/2\leftrightarrow -j = -l-1/2$, or equivalently $l\leftrightarrow-l-1$.
This symmetry implies that our boundary condition does not violates time reversal.
While in the vacuum the susceptibility functions actually become independent of $\Omega$, the $r$-dependence persists even in the vacuum due to the requirement to satisfy the boundary condition of the no-incoming flux condition in Eq.~\eqref{no_incoming}.
This behavior will be demonstrated in detail later using a model approach.

To investigate the $\Omega$-dependence of the susceptibility functions, we hereafter work within the rotating frame at finite temperatures.
To incorporate the finite temperature $T$ into the Dirac propagator, we adopt the imaginary-time formalism by introducing
the imaginary time $\tau$ through $t=-i\tau$.
Accordingly, the Matsubara propagator is defined as
\begin{equation}
\label{eq:Delta}
\Delta(x,x')
=T
 \sum_{n=-\infty}^\infty
 \int \frac{dp_z}{2\pi} 
\sum_{k=1}^\infty 
 \sum_{l=-\infty}^\infty
 e^{-i\omega_n (\tau -\tau^{\prime }) +i p_z (z-z')}\frac{\mathcal{S}_{l}(-i\omega_n, p_{l,k},p_z;\boldsymbol{x}_\perp,\boldsymbol{x}_\perp')}{\pi R^2 N_{l,k}^2}
 ,
\end{equation}
with $\omega_n = (2n+1 )\pi T$ and $\beta=1/T$.
In the subsequent sections, the imaginary-time formalism is employed with a focus on finite-temperature systems. Throughout this discussion, spacetime integrals are expressed with the following shorthand notation, $\int d^4x = \int_0^\beta d\tau \int d^3x$.

%\red{
One can employ alternative  boundary condition, such as the MIT boundary condition~\cite{Ambrus:2015lfr,Chernodub:2016kxh,Zhang:2020hct}.
While different boundary conditions generally lead to quantitative differences in the evaluation of physical quantities, the MIT boundary condition shares several common features with those under our boundary condition.
First, under the MIT boundary condition, the rotational effect on thermodynamic quantities at zero temperature, similar to ours. Also, at high temperatures, the boundary effect is negligible compared to the thermal wavelength, $1/T\ll R$~\cite{Ebihara:2016fwa}, whereas at low temperatures the boundary effect becomes significant.
We note that the above formulation is essentially unchanged under different boundary conditions, except for the momentum discretization. 
% To avoid such boundary effects, in this study, we focus on the high-temperature regime around the pseudocritical temperature.
%}

%%%%%%%%%%%%%%%%%%%%%%%%%%%%%%%%%%%%%%%%%%%%
\section{Susceptibility functions within an NJL model} \label{sus_NJL_model}
To explicitly compute the susceptibility functions under rotation, we adopt a specific model that includes interaction terms. In particular, we employ the NJL model with two light quark flavors, which is widely used in studies of hot QCD matter. The corresponding Lagrangian is given by
\begin{equation}
\mathcal{L} =
\bar q(i\slashed{D}-m_0) q
+{\cal L}_{\rm int},
\label{NJL_Lag}
\end{equation}
where
$q=(u,d)^T$ denotes the quark field with two degenerate flavors and $m_0$ represents the current quark mass $m_0=m_u=m_d$. The interaction term ${\cal L}_{\rm int}$ consists of the four-point quark interactions,
\begin{equation}
{\cal L}_{\rm int}
 =
\frac{g_s+g_d}{2}
\left[
(\bar q q )^2
+ 
\sum_{a=1}^3
(\bar q i\gamma_5  \tau^a q )^2
\right]
+
\frac{g_s-g_d}
{2}
\left[
\sum_{a=1}^3
(\bar q\tau^a q )^2
+ 
(\bar q i\gamma_5  q )^2
\right],
\end{equation}
where $\tau^{i=1,2,3}$ are Pauli matrices with $\tau^0={\bm 1}_{2\times 2}$, and $g_s$ and $g_d$ are the coupling constants.
The four-point interaction terms characterized by $g_s$ are invariant parts under the global $U(2)_L\times U(2)_R$ chiral symmetry. On the other hand, 
the interaction described by $g_d$ breaks the $U(1)_A$ symmetry and corresponds to `t Hooft-Kobayashi-Maskawa determinant term~\cite{Kobayashi:1970ji,Kobayashi:1971qz,tHooft:1976rip,tHooft:1976snw}. 

Before performing explicit calculations of the susceptibility function, we review the Ward-Takahashi (WT) identity associated with the chiral symmetry~\cite{GomezNicola:2013pgq,GomezNicola:2016ssy,GomezNicola:2017bhm}.
This identity serves as a theoretical constraint and a guiding principle for the evaluation of the susceptibility function within the present model analysis.

As a step toward deriving the WT identity, we define the thermal average of an operator $\mathcal{O}(y)$, as follows:
\begin{equation}
\langle 
 \mathcal{O}(y) 
\rangle=
\frac{1}{Z}
\int \mathcal{D}\bar q \mathcal{D} q\,
\mathcal{O}(y)
\exp\left[
\int d^4x \,\mathcal{L}(\bar{q},q)
\right],
\quad
Z = \int \mathcal{D}\bar q \mathcal{D} q\,
\exp\left[
\int d^4x^\prime \,\mathcal{L}(\bar{q},q)
\right].
\label{picon}
\end{equation}
Now we introduce the $SU(2)_A$-transformed integral variables as
\begin{equation}
\label{eq:su2A}
q' = (1+i\epsilon^i \gamma_5 \tau^i/2)q,
\quad
\bar{q}'
= \bar{q}(1+i\epsilon^i \gamma_5 \tau^i/2),
\end{equation}
where $\epsilon^i$ is a constant infinitesimal parameter.
For the pseudoscalar $\mathcal{O}(y)=\bar{q}(y)i\gamma_5\tau^i q
(y)$, the following relation holds:
\begin{equation}
\label{eq:ps-inv}
 \int \mathcal{D}\bar q' \mathcal{D} q'\,
\bar{q}'(y)i\gamma_5\tau^i q'(y)
\exp\left[
\int d^4x \,\mathcal{L}(\bar{q}',q')
\right]
= \int \mathcal{D}\bar q \mathcal{D} q\,
\bar{q}(y)i\gamma_5\tau^i q(y)
\exp\left[
\int d^4x \,\mathcal{L}(\bar{q},q)
\right]
\end{equation}
Under the transformation~\eqref{eq:su2A},  no nontrivial Jacobian emerges in the functional measure, but the Lagrangian and the pseudoscalar change as
\begin{equation}
 \mathcal{L}(\bar{q}',q')
 = \mathcal{L}(\bar{q},q) - m_0\epsilon^i \bar{q}(x)i\gamma_5\tau^i q(x),
 \quad
 \bar{q}'(x)i\gamma_5\tau^i q'(x)
 = \bar{q}(x)i\gamma_5\tau^i q(x) - \epsilon^i \bar{q}(x)q(x) .
\end{equation}
Plugging them into Eq.~\eqref{eq:ps-inv} and comparing the coefficients of $\epsilon^i$ in both sides,
we can obtain the desired WT identity,
\begin{equation}
\langle \bar q q \rangle (x)
= - m_0 \chi_\pi(x),
\label{WTISU2}
\end{equation}
with $\chi_\pi$ being the pion susceptibility,
\begin{equation}
\chi_\pi (x) = 
\int d^4x^\prime
\langle
T_\tau
\bar q(x)i\gamma_5\tau^i q (x)
\bar q(x')i\gamma_5\tau^i q (x')
\rangle.
\end{equation}
Note that the isospin index $i$ is not summed over and the results for $i=1,2,3$ are identical due to the isospin symmetry of the NJL model.
Through the WT identity in Eq.~\eqref{WTISU2}
the pion susceptibility is directly related to the quark condensate.
A crucial remark here is that the above argument is irrelevant to coordinates and boundary conditions, as we have not specified it above. Hence, the WT identity~\eqref{WTISU2} should be respected even in our rotating system.

In the NJL model, the quark condensate acquires a nonzero value, indicating the occurrence of spontaneous chiral symmetry breaking. To incorporate this feature into the model analysis, we impose the mean-field approximation. Within this framework, the interaction terms take the following form,
\begin{equation}
{\cal L}_{\rm int} \to 
(g_s+g_d)\langle 
 \bar qq
 \rangle\bar qq
 -\frac{g_s+g_d}{2}
 \langle 
 \bar qq
 \rangle^2 + {\cal L}_{\rm int}.
\end{equation}
Here, the interaction term $ {\cal L}_{\rm int}$ is interpreted as a fluctuation around the vacuum in which spontaneous chiral symmetry breaking occurs.
Under the mean-field approximation, the effective action and the partition function are described by $\Gamma= \ln Z$ with
\begin{equation}
\begin{split}
Z =
\int \mathcal{D}\bar q \mathcal{D} q
\exp\left[
\int d^4x \,\mathcal{L}_\mathrm{mean}
\right],
\qquad
{\mathcal{L}}_{\rm mean}
=
\bar q\Bigl[i\slashed{D}-M(r)\Bigl] q
- \frac{[M(r)-m_0]^2}{2(g_s + g_d)}
% - \frac{[M(r)-m_0]^2}{2(g_s + g_D)}
+{\cal L}_{\rm int}
,
\end{split}
\label{eq:Lmean}
\end{equation}
where the three-dimensional spatial integral is given by $\int d^3x = \int_0^R dr\,r \int_0^{2\pi} d\theta \int_{-\infty}^\infty dz$, and $M$ denotes the dynamical quark mass, $M= m_0 - (g_s+g_d)\langle \bar q q\rangle$.
In the rotating cylindrical system,
the quark condensate depends on the radial coordinate $r$ as a consequence of the translational noninvariance, and so does the dynamical mass $M=M(r)$.
This effective Lagrangian implies that the quark Matsubara propagator $\Delta_\mathrm{q}(x,x')=\langle 0|T_\tau q(x) \bar q(x')|0\rangle$ with the imaginary time ordering $T_\tau$ is the solution of the following equation:
\begin{equation}
\label{eq:eom_S}
\left[i\slashed{D}_x-M(r)\right]
\Delta_\mathrm{q}(x,x';M(r)) 
=-\delta(\tau-\tau')\delta^{(3)}(\boldsymbol{x}-\boldsymbol{x}').
\end{equation}
While in general this cannot be solved without specifying the inhomogeneity of the mass, under the local density approximation the solution is represented as the same form as the free theory, up to $O(\partial_r M/M^2)$.
That is, the propagator is given by Eq.~\eqref{eq:Delta} but with the replacement of the mass, as follows:
\begin{equation}
\label{eq:Sq}
 \Delta_\mathrm{q}(x,x';M(r)) 
 = \Delta(x,x')\Big|_{m\to M(r)}\,, 
\end{equation}
where we implicitly incorporate the isospin degrees of freedom.

For later convenience, let us here show an important relation of the above propagator.
Using the property of the time ordering in $\Delta_\mathrm{q}(x,x')=\langle 0|T_\tau q(x) \bar q(x')|0\rangle$, we show that
\begin{equation}
\Delta_\mathrm{q}(x^\prime,x;M(r'))
=-\gamma^0 \Delta_\mathrm{q}^\dagger(x,x^\prime;M(r))\gamma^0.
\end{equation}
Plugging the solution~\eqref{eq:Sq} into this relation, we can arrive at
\begin{equation}
\label{eq:sameM}
 \Delta_\mathrm{q}(x',x;M(r'))
 = \Delta_\mathrm{q}(x',x;M(r)).
\end{equation}
This relation is advantageous in evaluating two-point correlations of quark-bilinears, which we analyze in the following sections.
Thanks to Eq.~\eqref{eq:sameM}, such correlators involving the product of two propagators are described with only one mass, enabling us to employ the same computational manner as that of the free theory.

%%%%%%%%%%%%%%%%%%%%
\subsection{Meson susceptibilities}
In this subsection, we present the susceptibility functions using the quark (Matsubara) propagator based on the Fourier-Bessel basis, under both the mean-field approximation and the local density approximation. 
In a similar manner to that in the subsection~\ref{sus_def}, we introduce the spurion term
\begin{equation}
\begin{split}
{\cal L}_{\rm sp} &=
\zeta_{\sigma} {\cal O}_\sigma
+\zeta_{\pi} {\cal O}_{\pi}
+\zeta_{\delta} {\cal O}_\delta
+
\zeta_{\eta} {\cal O}_\eta,
\end{split}
\label{sp_term}
\end{equation}
with
\begin{equation}
{\cal O}_\sigma = \bar q q,\quad
{\cal O}_{\pi} = \bar q  i\gamma_5\tau^3 q,
\quad
{\cal O}_\eta = \bar q  i\gamma_5 q, 
\quad 
{\cal O}_\delta = \bar q  \tau^3 q.
\label{operators}
\end{equation}
Note that we solely focus on the third isospin component of the quark bilinears corresponding to the pion and delta, as isospin symmetry guarantees equivalent results for the first and second components.
By including it in the NJL Lagrangian in Eq.~\eqref{NJL_Lag}, the susceptibility functions can be defined as
\begin{equation}
\begin{split}
\int d^4x
\chi_\alpha  (x)
&=
\frac{\partial^2\Gamma}{\partial \zeta_{\alpha}^2 
 }
\Biggl{|}_{\zeta_\alpha =0}
=
\int d^4x
\int d^4x' \langle T_\tau
{\cal O}_\alpha (x)\, 
{\cal O}_\alpha (x')
\rangle,
\qquad
\alpha = \pi, \eta,\sigma,\delta. 
\label{meson_sus_def}
\end{split}
\end{equation}
Indeed, the following calculation demonstrates that the meson susceptibility functions depend on the radial coordinate $r$ and the dynamical quark mass $M(r)$, reflecting the violation of translational symmetry.

As the first demonstration, we present the evaluation of the pion susceptibility~$\chi_\pi$:
\begin{equation}
\begin{split}
&
\int d^4x\,
\chi_\pi \left(r;M(r)\right)
=
\int d^4x
\int d^4x' 
\left\langle 
T_\tau{\cal O}_{\pi}(x)\, {\cal O}_{\pi}(x') 
\exp\left[
\int d^4x_1\frac{G_\pi}{2}(\bar q i\gamma_5\tau^k q)^2
\right]
\right\rangle_0,
\label{pion_sus}
\end{split}
\end{equation}
with $G_\pi=g_s+g_d$ and $\langle\cdots\rangle_0$ representing the vacuum expectation value under the mean-field approximation without the fluctuating interaction term ${\cal L}_{\rm int}$. 
Furthermore, we have extracted the relevant four-point quark interaction term for the pion susceptibility, characterized by $G_\pi$. To proceed with the evaluation of $\chi_\pi$, we will perform an expansion in terms of $G_\pi$.
The leading term of the $G_\pi$ expansion in Eq.~\eqref{pion_sus} is identified as the one-loop result of the pion susceptibility $\chi_\pi^{\rm (1-loop)}\left(r;M(r)\right)$, which is written as follows:
\begin{equation}
\begin{split}
&\int d^4x
\chi_\pi^{\rm (1-loop)}\left(r;M(r)\right) \\
&=
\int d^4x
\int d^4x' \langle 
T_\tau{\cal O}_{\pi}(x)\, {\cal O}_{\pi}(x') 
\rangle_0\\
&=
-
N_cN_f
\int d^4x
\int d^4 x^\prime
{\rm tr}
\left[
\Delta_\mathrm{q}\left(x,x^\prime;M(r)\right)
i\gamma_5 
\Delta_\mathrm{q} \left(x^\prime,x;M(r^\prime)\right)
i\gamma_5 
\right],
\end{split}
\label{1loop}
\end{equation}
where $N_c=3$ and $N_f=2$ denote the number of colors and flavors, respectively.

To proceed with the evaluation, we mention two important remarks.
The first one is about the relation between the integrands in both sides.
While in translational invariant systems, this equation would provides the equality between two integrands, in the present case the difference between them are, without loss of generality, described by an $r$-dependent redundant $\alpha_1(r)$, whose radial integral vanishes.
Another is that thank to the relation~\eqref{eq:sameM}, the integration over $x'$ as the usual fermionic one-loop diagram with the mass $M(r)$.
Therefore, we obtain
\begin{equation}
\chi_{\pi}^{({\rm 1-loop})}
\left(r;M(r)\right)
=
N_c N_f \int\frac{dp_z}{2\pi}
\sum_{l,k}
\frac{J_{l}(p_{l,k} r)^2 + J_{l+1}(p_{l,k} r)^2 }{\pi R^2 N_{l,k}^2}
\frac{f(p_z;\Lambda)
-2 n_F(E)}{E}
 +\alpha_1(r)
\label{one_loop_pisus}
\end{equation}
with $E = \sqrt{p_{l,k}^2 +p_z^2 +M(r)^2}$ and
the Fermi distribution function $n_F(E) = [e^{\beta\{E -\Omega(l+1/2)\}} +1]^{-1}$.
This quark one-loop calculation involves an ultraviolet (UV) divergence. To regularize this UV divergence, we introduce the following smooth regulator function into the effective Lagrangian,
\begin{equation}
\label{eq:f}
f(p_z;\Lambda)= 
\frac{\Lambda^{10}}{
\Lambda^{10} + 
\left(
\sqrt{ p_{l,k}^2 +p_z^2}
\right)^{10}
},
\end{equation}
with $\Lambda$ being the UV cutoff.
As for the rotational effect, its explicit dependence appears only in the Fermi distribution function.
At zero temperature, however, the distribution function vanishes due to the infrared energy gap and the causality constraint, at least for the present discretized momentum~\eqref{eq:pperp}.
In other words, there is no visible rotational effect on thermodynamic quantities, as shown in Ref.~\cite{Ebihara:2016fwa}.
The same feature is also found for a different boundary condition~\cite{Chernodub:2016kxh} and~for a rotating sphere~\cite{Zhang:2020hct,Mameda:2023sst}.
 
Similarly to the leading order computation, higher-order terms are evaluated by introducing redundant parameters in each order.
Eventually, the pion susceptibility in the rotating frame can be expressed as 
\begin{equation}
\begin{split}
\int d^4x
\chi_\pi (r;M(r))
&=
\int d^4x
\left[
\left\{
\chi_\pi^{\rm (1-loop)}
+
G_\pi
\left(
\chi_\pi^{\rm (1-loop)}
\right)^2
+
G_\pi^2
\left(
\chi_\pi^{\rm (1-loop)}
\right)^3
+\cdots
\right\}
+ \mathcal{R}
\right]\\
&=
\int d^4x
\frac{\chi_\pi^{({\rm 1-loop})} }{1-G_\pi \chi_\pi^{({\rm 1-loop})} }
+ \int d^4x \mathcal{R}
,
\end{split}
\end{equation}
where $\mathcal{R}=\mathcal{R}(\alpha_1(r),\alpha_2(r),\cdots)$ is the total redundant contribution determined by $\alpha_n(r)$ being the parameter emerging in the $O(G_\pi^n)$ computation.
The pion susceptibility is now represented in the bubble ring diagram form, as in the vacuum case~\cite{Hatsuda:1994pi}, accompanied by $\mathcal{R}$.
Indeed, when attempting to satisfy the WT identity in Eq.~\eqref{WTISU2} under both the mean-field approximation and the local density approximation, the redundant term must vanish. That is, by respecting the WT identity, the pion susceptibility is given by the following expression:
\begin{equation}
\begin{split}
\chi_\pi\left(r;M(r)\right)
=
\frac{\chi_\pi^{({\rm 1-loop})} }{1 - G_\pi \chi_\pi^{({\rm 1-loop})} }.
\end{split}
\label{chipi}
\end{equation}
A detailed verification of whether this pion susceptibility satisfies the WT identity within the current NJL model analysis will be examined at the beginning of section~\ref{numerical_sus}.

Besides, using the expression of the pion susceptibility in Eq.~\eqref{chipi} as a basis, we proceed to derive the explicit forms of the susceptibility functions for other meson channels. 
Through the $SU(2)_A$ and $U(1)_A$ transformations, the quark bilinears in Eq.~\eqref{operators} are transformed into different quark bilinears as%
~\footnote{
These transformations are performed in the context of bilinear field operators $\mathcal{O}_\alpha$, but the same results are also derived in the path integral formalism.
That is, WT identities similar to Eq.~\eqref{WTISU2} are obtained through the variable transformation, and these identities requires Eq.~\eqref{sus_NJL}. }
\begin{equation}
\begin{split}
{\cal O}_{\pi} \overset{SU(2)_A}{\rightarrow} {\cal O}_\sigma \overset{U(1)_A}{\rightarrow} {\cal O}_\eta \overset{SU(2)_A}{\rightarrow} {\cal O}_\delta.
\end{split}
\end{equation}
By taking into account the transformations with the result of the pion susceptibility in Eq.~\eqref{chipi}, 
the meson susceptibilities, including those of other mesonic channels, are described as follows:
\begin{equation}
\begin{split}
\chi_\alpha \left(r;M(r)\right) 
&=\frac{\int d^4x' \langle 
T_\tau
 {\cal O}_\alpha
(x)\, 
 {\cal O}_\alpha
(x') 
\rangle_0}{
1-G_\alpha\int d^4x' \langle 
T_\tau
 {\cal O}_\alpha
(x)\, 
 {\cal O}_\alpha
(x') 
\rangle_0
}
\\
&=
\frac{\chi_\alpha^{({\rm 1-loop})} }{1
-
G_\alpha \chi_\alpha^{({\rm 1-loop})} }
,
\qquad
\alpha = \pi, \eta,\sigma,\delta,
\label{sus_NJL}
\end{split}
\end{equation}
where the effective couplings are 
\begin{eqnarray}
G_\sigma = G_\pi,\quad
G_\eta = G_\delta=g_s-g_d.
\end{eqnarray}
The one-loop results are given by $\chi_{\eta}^{({\rm 1-loop})} = \chi_{\pi}^{({\rm 1-loop})}$ and $\chi_{\sigma}^{({\rm 1-loop})}=\chi_{\delta}^{({\rm 1-loop})}$ with
\begin{equation}
\begin{split}
\chi_{\sigma}^{({\rm 1-loop})}\left(r;M(r)\right)
&=
N_c N_f \int\frac{dp_z}{2\pi}
\sum_{l,k}
\frac{J_{l}(p_{l,k} r)^2 + J_{l+1}(p_{l,k} r)^2 }{\pi R^2 N_{l,k}^2}\\
&\quad\times
\left[
\frac{ p_{l,k}^2+p_z^2 }{ 
E^3 
}
\Bigl\{
f(p_z;\Lambda)
-
2n_F(E)
\Bigr\}
-
\frac{2M^2}{TE^2}
n_F'(E)
\right],
\end{split}
\label{one_loop_analytical}
\end{equation}
where $n_F'(E) = d n_F/dE$.
As with the one-loop result for the pion susceptibility, the resumed meson susceptibilities $\chi_\alpha$ also become independent of rotation $\Omega$ at zero temperature. The rotational effect at finite temperature will be specifically analyzed through numerical calculations in the next section.

%%%%%%%%%%%%%%%%%%%%%%%%%%%%%%%%%%%%%%%%%%%%
\subsection{Other susceptibility functions}
In this subsection, we also analytically examine other susceptibility functions that can be described within the NJL model framework under rotation, including the topological susceptibility, baryon number susceptibility and moment of inertia.

%%%%%%%%%%%%%%%%%%%%%%%%%%%%%%%%%%%%%%%%%%%%
\subsubsection{Topological susceptibility}
To define the topological susceptibility in the underlying QCD theory, $\theta$-term 
is introduced into the QCD Lagrangian as
\begin{eqnarray}
{\cal L}_{\theta}=\theta Q,
\quad
Q=\frac{g^2}{64\pi^2}\epsilon^{\mu\nu\rho\sigma} G_{\mu\nu}^a G_{\rho\sigma}^a,
\end{eqnarray}
where $\theta$ represents the $\theta$ parameter treated as an external field,
$Q$ denotes the topological operator with $g$ being the QCD coupling constant and $G_{\mu\nu}^a$ being the field strength of gluons.
Taking the second derivative
of the QCD effective action $\Gamma_{\rm QCD}$ with respect to the $\theta$ parameter yields the topological susceptibility,
\begin{equation}
\int d^4x
\chi_{\rm top} (x)
=
\frac{\partial^2 \Gamma_{\rm QCD} }{\partial \theta^2} 
\Biggl{|}_{\theta=0}.
\label{def_top}
\end{equation}
By further evaluating the right-hand side in Eq.~\eqref{def_top} with the $\theta$-term, the topological susceptibility
is expressed by the two-point correlation function of the topological operator, 
\begin{equation}
\chi_{\rm top} = 
\int d^4x^\prime
\langle T_\tau Q(x)
Q(x')
\rangle.
\end{equation}
In the NJL model without rotation, this is evaluated as a constant form~\cite{GomezNicola:2016ssy,GomezNicola:2017bhm,GomezNicola:2019myi,Kawaguchi:2020qvg,Cui:2021bqf,Cui:2022vsr}: $\chi_{\rm top} = \frac{m_0}{4} \langle \bar q q \rangle+\frac{m_0^2}{4} \chi_\eta$.
% \int d^4x' \langle T\bar q i\gamma_5 q (x)\bar q i\gamma_5 q (x')\rangle.
Under rotation, it becomes the radial-coordinate dependent form, as follows:
\begin{equation}
\chi_{\rm top}\left(r;M(r)\right)
=
\frac{m_0^2}{4}
\left[
-\chi_\pi\left(r;M(r)\right)
+\chi_\eta\left(r;M(r)\right)
\right],
\label{chi_top_alt}
\end{equation}
where we have used the WT identity~\eqref{WTISU2}. 

% One might think that this expression may not be directly manageable within the effective model analysis, as the gluon degrees of freedom are not explicitly included.
% However, the topological susceptibility is a well-defined quantity even within low-energy effective theories and models.
We should here emphasize that the topological susceptibility is a well-defined quantity even within low-energy effective theories and models which have no explicit gluon degrees of freedom.
This is because the full QCD involving the $\theta$ term is mathematically equivalent to that without the $\theta$ term but with the quark mass $m\,e^{i\gamma_5\theta/N_f}$, which was pointed out in Ref.~\cite{Peccei:1977hh}.
This equivalence is the basis in constructing the Di-Vecchia--Veneziano model~\cite{DiVecchia:1980ve},
% , based on the chiral perturbation theory (ChPT).
which is an effective model with finite $\theta$ and provides a dynamical mechanism underlying in the Veneziano-Witten formula~\cite{Witten:1979vv,Veneziano:1979ec}.
% , with which we understand that the large mass of $\eta'$ is due to the topological susceptibility in pure Yang-Mills theory.
% One of the guiding principles on constructing the DVV model is the above connection between the topological charge and axial anomaly through the $\mathrm{U(1)}_\mathrm{A}$ transformation (for the detailed procedure, see, e.g., Ref.~\cite{GomezNicola:2016ssy,Kawaguchi:2020qvg})
% :
% \begin{equation}
% -m_0\bar qq + \theta Q\to 
% -m_0\bar q
% \exp\left[
% i\gamma_5 \theta/2
% \right]q.
% \label{theta_mass}
% \end{equation}
In the same manner, we can introduce $\theta$ in other works (see Ref.~\cite{Leutwyler:1992yt} for example), including the NJL model in Refs.~\cite{Boer:2008ct}.
These effective models lead to the topological susceptibility consistent with that derived from the full QCD with the mass term $\bar{q}(m\, e^{i \gamma_5\theta/N_f})q$~\cite{GomezNicola:2016ssy,GomezNicola:2017bhm,GomezNicola:2019myi,Kawaguchi:2020qvg,Cui:2021bqf,Cui:2022vsr}, and indeed shows good agreements with lattice QCD~\cite{Borsanyi:2016ksw,Kawaguchi:2020qvg,Cui:2021bqf}
% \begin{equation}
% \chi_{\rm top}
% =
% \frac{m_0}{4}
% \langle \bar q q \rangle
% +\frac{m_0^2}{4} \chi_\eta
% %\int d^4x' \langle T\bar q i\gamma_5 q (x)\bar q i\gamma_5 q (x')\rangle
% \label{chitop_flat}
% .
% \end{equation}
% Importantly, one can show that the same expression of $\chi_\mathrm{top}$ is also derived from the full QCD with the mass term $\bar{q}(m\, e^{i \gamma_5\theta/N_f})q$.
% Based on this analytical expression,
\footnote{Indeed, the model calculations of the topological susceptibility show good agreements with the lattice QCD data in flat spacetime without rotation~\cite{Borsanyi:2016ksw,Kawaguchi:2020qvg,Cui:2021bqf}, as follows:
\begin{center}
\begin{tabular}{ll}
2+1+1 flavor lattice QCD (at $T=0$)~\cite{Borsanyi:2016ksw}: &
$\chi_{\rm top}=0.0245(24)_{\rm stat}(03)_{\rm flow}(12)_{\rm cont}/{\rm fm}^4$ \\
3 flavor ChPT~\cite{Borsanyi:2016ksw} (see also Ref.~\cite{Leutwyler:1992yt}): &
$\chi_{\text{top}} = 0.0224/{\rm fm}^4$ \\
3 flavor LSM~\cite{Kawaguchi:2020qvg}: &
$\chi_{\rm top}=0.0263/{\rm fm}^4$ \\
3 flavor NJL~\cite{Cui:2021bqf}: &
$\chi_{\rm top}=0.025/{\rm fm}^4$, \\
\end{tabular}
\end{center}
%%%
% \begin{equation*}
% \begin{split}
% \text{2+1+1 flavor lattice QCD (at $T=0$)}~\cite{Borsanyi:2016ksw} :\quad
% \chi_{\rm top}&=0.0245(24)_{\rm stat}(03)_{\rm flow}(12)_{\rm cont}/{\rm fm}^4 \\
% \text{3 flavor ChPT~\cite{Borsanyi:2016ksw}}
% \text{ (see also Ref.~\cite{Leutwyler:1992yt})}
% : \quad
% \chi_{\text{top}} 
% &= 0.0224/{\rm fm}^4 \\
% % {\rm LSM\,w/\,3\,flavor
% \text{3 flavor LSM~\cite{Kawaguchi:2020qvg}}:\quad {\chi}_{\rm top}
% &=0.0263/{\rm fm}^4\\
% \text{3 flavor NJL~\cite{Cui:2021bqf}}:
% \quad {\chi}_{\rm top}
% &=0.025/{\rm fm}^4,
%\end{split}
%\end{equation*}
 where LSM stands for the linear sigma model.}.
% Let us now consider the topological susceptibility in the rotating matter.
Since the $\mathrm{U(1)}_\mathrm{A}$ transformation is intact even under rotation,
% in the rotational cylindrical frame, 
the topological susceptibility~\eqref{chi_top_alt} takes a form similar to that in the nonrotating case, except for the $r-$ and $\Omega$-dependence.
\subsubsection{Baryon number susceptibility}
Within the framework of the NJL model, the baryon number susceptibility can also be evaluated. 
In order to derive the baryon number susceptibility $\chi_B$, we temporarily introduce the chemical potential associated with the baryon charge density operator, $j_B = \bar q \gamma^0 q$, into the covariant derivative:
$i\slashed{D}\to i\slashed{D}+ \mu\gamma^0$.
The physical quantity of current interest is then obtained by taking the second derivative of the NJL effective action with respect to $\mu$,
\begin{equation}
\begin{split}
\int d^4x
\chi_{B}(x)
&=
\frac{\partial^2 \Gamma}{\partial \mu^2}
\Biggl{|}_{\mu=0}
=
\int d^4x
\int d^4x' \langle 
T_\tau
j_B(x)\, 
j_B(x') 
\rangle.
\end{split}
\end{equation}
In the NJL model, by using the quark propagate in the Fourier-Bessel basis, the baryon number susceptibility is evaluated at the quark one-loop level as
\begin{equation}
\chi_{B} \left(r;M(r)\right)
=
N_c N_f \int\frac{dp_z}{2\pi}
\sum_{l,k}
\frac{J_{l}(p_{l,k} r)^2 + J_{l+1}(p_{l,k} r)^2 }{\pi R^2 N_{l,k}^2}
\left\{
-2 n_F'(E)
\right\}.
\label{baryon_sus}
\end{equation}
This shows that the baryon number susceptibility does not have the UV divergence, but rather contains only the thermal correction associated with the Fermi distribution function.
Therefore, in the vacuum ($T=0$), the baryon number susceptibility vanishes even in the presence of the finite $\Omega$.
The $\Omega$-dependence at finite temperature will also be evaluated numerically in the next section.

%%%%%%%%%%%%%%%%%%%%%%%%%%%%%%%
\subsubsection{Moment of inertia}
By treating the angular velocity $\Omega$ as an external field, the quark bilinear fields that couple to $\Omega$ in the action can be identified with the total angular momentum operator,
$j_\Omega = \bar q
\gamma^0 \left(
-i\partial_\theta
+\sigma^{12}/2\right)q$.
Based on this operator, one can define a susceptibility function evaluated at $\Omega = 0$, which corresponds to the two-point correlation function of $j_\Omega$. Physically, this susceptibility function represents the moment of inertia of the finite cylinder filled with quark matter:
\begin{equation}
\begin{split}
\int d^4x
\chi_{\Omega}(x) &
=
\frac{\partial^2 \Gamma}{\partial \Omega^2}
\Biggl{|}_{\Omega=0}
=
\lim_{\Omega\to 0}\left[
\int d^4x
\int d^4x' \langle
T_\tau
j_\Omega(x)\, 
j_\Omega(x') 
\rangle
\right].
\label{def_MoI}
\end{split}
\end{equation}
Similar to the baryon number susceptibility, the moment of inertia is also evaluated at the quark one-loop level as
\begin{equation}
\chi_{\Omega}\left(r;M(r)\right) =
N_c N_f \int\frac{dp_z}{2\pi}
\sum_{l,k}
\frac{J_{l}(p_{l,k} r)^2 + J_{l+1}(p_{l,k} r)^2 }{\pi R^2 N_{l,k}^2} 
\left\{
-2 n_F'(E)
\right\}
\left(l+\frac{1}{2}\right)^2.
\label{minertia}
\end{equation}
This analytic expression closely resembles that of the baryon number susceptibility with an additional factor corresponding to the square of the total angular momentum $j^2=(l+1/2)^2$.
This resemblance comes from the analogy in the finite density effect and the rotational effect.
Moreover, if the system were invariant under the radial translation, only the $j=\pm 1/2$ modes would survive, implying $\chi_\Omega = (1/2)^2 \chi_B|_{\Omega \to 0}$.

%%%%%%%%%%%%%%%%%%%%%%%%%%%%%%%%%%%%%%%%%%%%
\section{Numerical evaluation of susceptibilities} \label{numerical_sus}
In this section, we numerically compute susceptibility functions in a rotating cylindrical frame using the NJL model. As a preliminary step in the evaluation, we first determine the dynamical quark mass by solving the stationary condition of the NJL effective action:
\begin{equation}
\frac{\delta\, \Gamma\left(
M
\right)}{\delta M}=0.
\end{equation}
Within the mean-field approximation, the NJL effective action is obtained by integrating out the quark field at the one-loop level in the generating functional,
\begin{equation}
\begin{split}
\Gamma \left(
M
\right)T &=
-\int d^3x
\frac{(M-m_0)^2}{2(g_s + g_D)} 
+
N_c N_f\int d^3x
\int \frac{d p_z}{2\pi} 
\sum_{l,k}
\frac{J_l(p_{l,k} r)^2+J_{l+1}(p_{l,k} r)^2}{\pi R^2 N_{l,k}^2}\\
&\quad\times
\left[
E  
f(p_z;\Lambda)
+
2T\ln
\left(
1+
e^{-\beta \left\{
E
-\Omega\left(l + \frac{1}{2}\right)
\right\}}
\right)
\right].
\label{effective_action}
\end{split}
\end{equation}
By solving the stationary condition, we determine the dynamical quark mass,
\begin{equation}
M
=m_0 + M (g_s+g_d)
N_c N_f \int\frac{dp_z}{2\pi}
\sum_{l,k}
\frac{J_{l}(p_{l,k} r)^2 + J_{l+1}(p_{l,k} r)^2 }{\pi R^2 N_{l,k}^2}
\frac{f^{(l,k)}(p_z;\Lambda)
- 2 n_F(E)}{E}. 
\label{dynmical_m}
\end{equation}
This surely shows that the dynamical quark mass acquires an $r$-dependence.

Using the dynamical quark mass in Eq.~\eqref{dynmical_m}, one can verify the validity of the WT identity in Eq.~\eqref{WTISU2} within the NJL model under rotation.
From the dynamical quark mass, the quark condensate can be expressed as
\begin{eqnarray}
\langle \bar qq\rangle = - M
N_c N_f \int\frac{dp_z}{2\pi}
\sum_{l,k}
\frac{J_{l}(p_{l,k} r)^2 + J_{l+1}(p_{l,k} r)^2 }{\pi R^2 N_{l,k}^2}
\frac{f(p_z;\Lambda)
- 2 n_F(E)}{E}.
\end{eqnarray}
Substituting this into the pion susceptibility in Eq.~\eqref{chipi}, one finds that the WT identity is satisfied:
\begin{equation}
\begin{split}
\chi_\pi
&=\frac{-\langle \bar qq\rangle}{
M+G_\pi
\langle \bar qq\rangle
}=
\frac{-\langle \bar qq\rangle}{m_0}.
\end{split}
\end{equation}

When formulating the quark propagator within the local density approximation, we have analytically neglected the $O(\partial_r M/M^2)$ term in  Eq.~\eqref{eq:Sq}. Therefore, our numerical analysis is valid provided that the $r$-dependent quark mass satisfies the following condition:
\begin{eqnarray}
\frac{|\partial_r M|}{M^2} \ll 1.
\end{eqnarray}
To examine this condition, we implement a numerical calculation with the following model parameters:
$\Lambda = 681.38\,{\rm MeV}$,
$m_0=4.552\, {\rm MeV}$,
$g_0 =1.860/\Lambda^2$,
$g_s=2(1-c)g_0$,
$g_d=2cg_0$ and 
$c=0.2$. 
In this parameter setting, the dynamical quark mass is evaluated as $M_{\rm vac}= 286{\rm MeV}$ at the vacuum.
Furthermore, in the absence of rotation and boundary conditions, the NJL model provides the chiral crossover, in which the pseudocritical temperature is estimated to be $T_{\rm pc}=168\,{\rm MeV}$. 
Hereafter, when considering the rotating cylindrical frame, we set the radius of a finite-size cylinder to $R=30/\Lambda$, and choose the angular velocity as $\Omega=0.5/R$.

\begin{figure}
\begin{tabular}{cc}
\begin{minipage}{0.5\hsize}
\begin{center}
    \includegraphics[width=0.95\columnwidth]{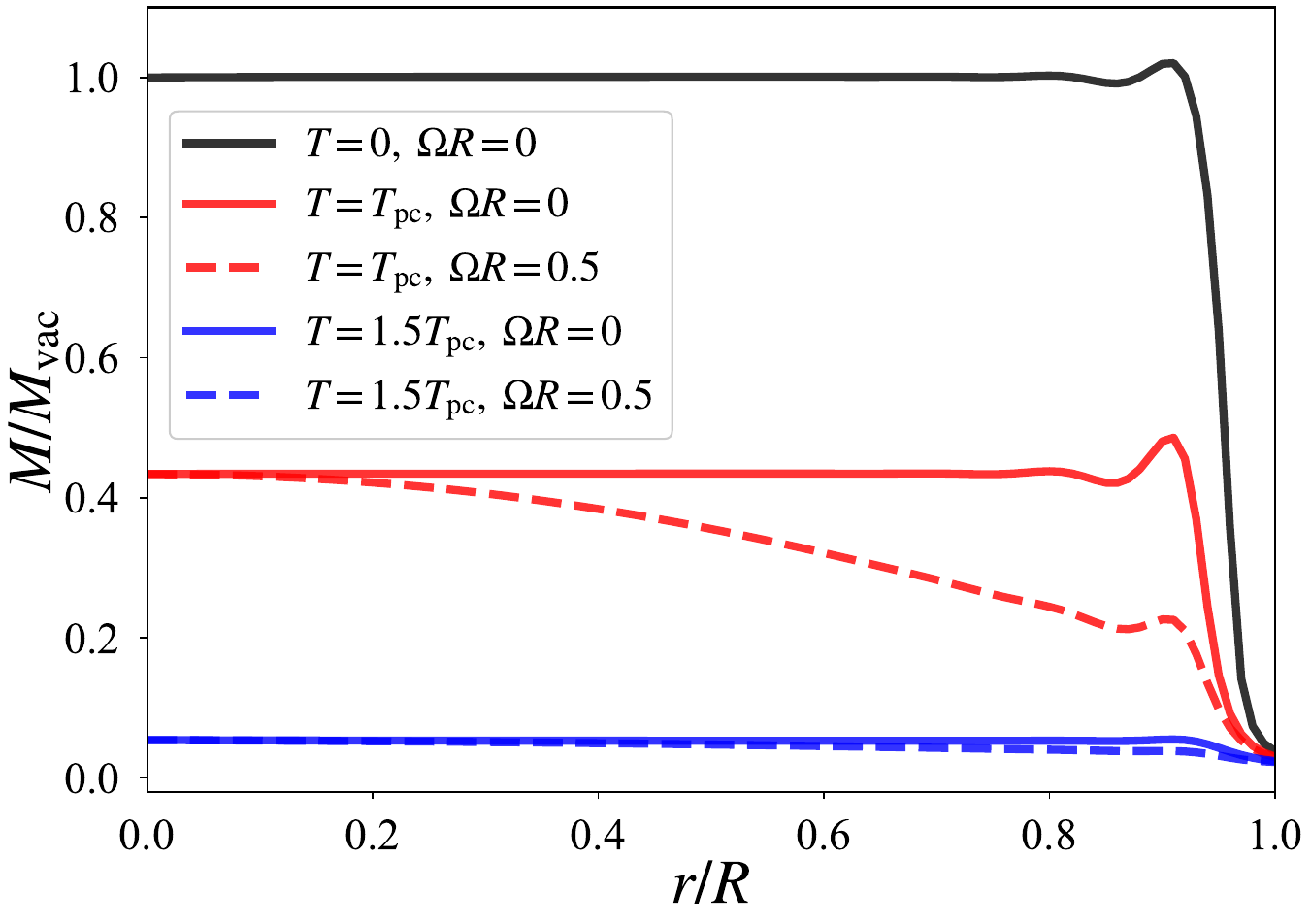}
    \subfigure{(a)}
\end{center}
\end{minipage}
\begin{minipage}{0.5\hsize}
\begin{center}
    \includegraphics[width=0.95\columnwidth]{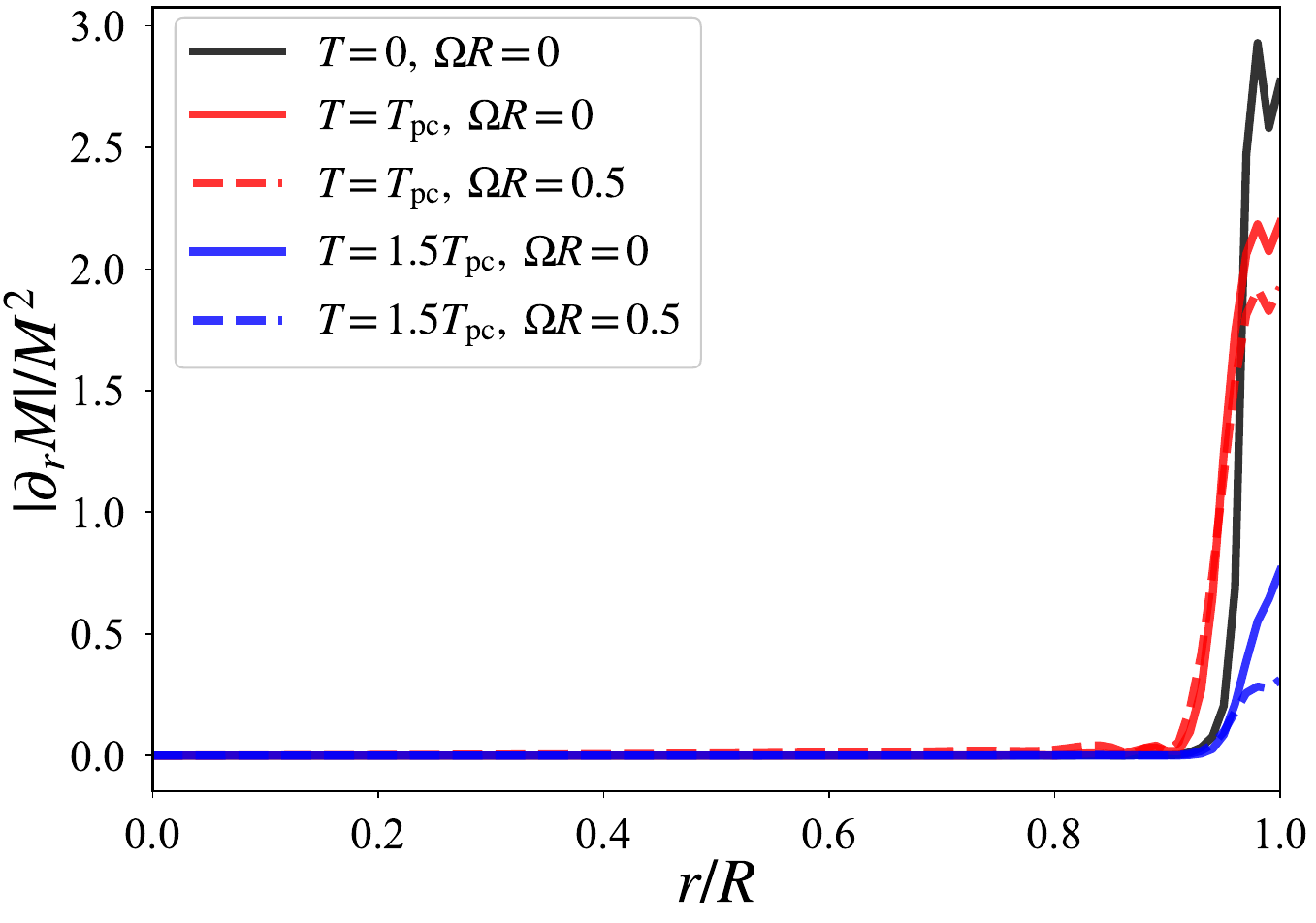}
    \subfigure{(b)}
\end{center}
\end{minipage}
\end{tabular}
\caption{
The rotational effect on the dynamical quark mass normalized by its vacuum value $M_{\rm vac}$: (a)
the $r$-dependence of the dynamical quark mass at fixed temperatures, and 
(b) the numerical validation of
the local density approximation.
}
\label{dynamical_mass_rotate}
\end{figure}

As we mentioned below Eq.~\eqref{eq:f}, at $T=0$ the dynamical quark mass is unaffected by finite $\Omega$ effects, though it still exhibits $r$-dependence. To numerically examine the $r$-dependence of the dynamical quark mass in the cylindrical frame at fixed temperatures, 
we present panel~(a) of Fig.~\ref{dynamical_mass_rotate}. At $T=0$, the dynamical quark mass remains constant up to approximately $r=0.85\,R$ but begins to decrease sharply near the boundary to satisfy the boundary condition. Moving on to finite temperature systems, the dynamical quark mass is affected by the rotational effect even for $r<0.85R$, and it decreases with increasing $r$. 

To confirm the validity of the local density approximation, we plot $|\partial_r M|/M^2$ in panel~(b).
This result shows that $|\partial_r M|/M^2 \ll 1$ is satisfied for approximately $r < 0.9\,R$ even at high temperatures across the pseudocritical temperature with the finite $\Omega$. This implies that the local density approximation is valid up to $r = 0.9\,R$ even when the rotational effect is included at the high temperatures. However, in the region near the boundary for $0.9\,R < r < R$, $|\partial_r M|/M^2$ increases sharply, and the validity of the local density approximation becomes unreliable. Based on this evaluation, our analysis in the following subsections will focus on the region $r < 0.9\,R$.

\begin{figure} %htbp
\begin{center}
    \includegraphics[width=0.47\columnwidth]{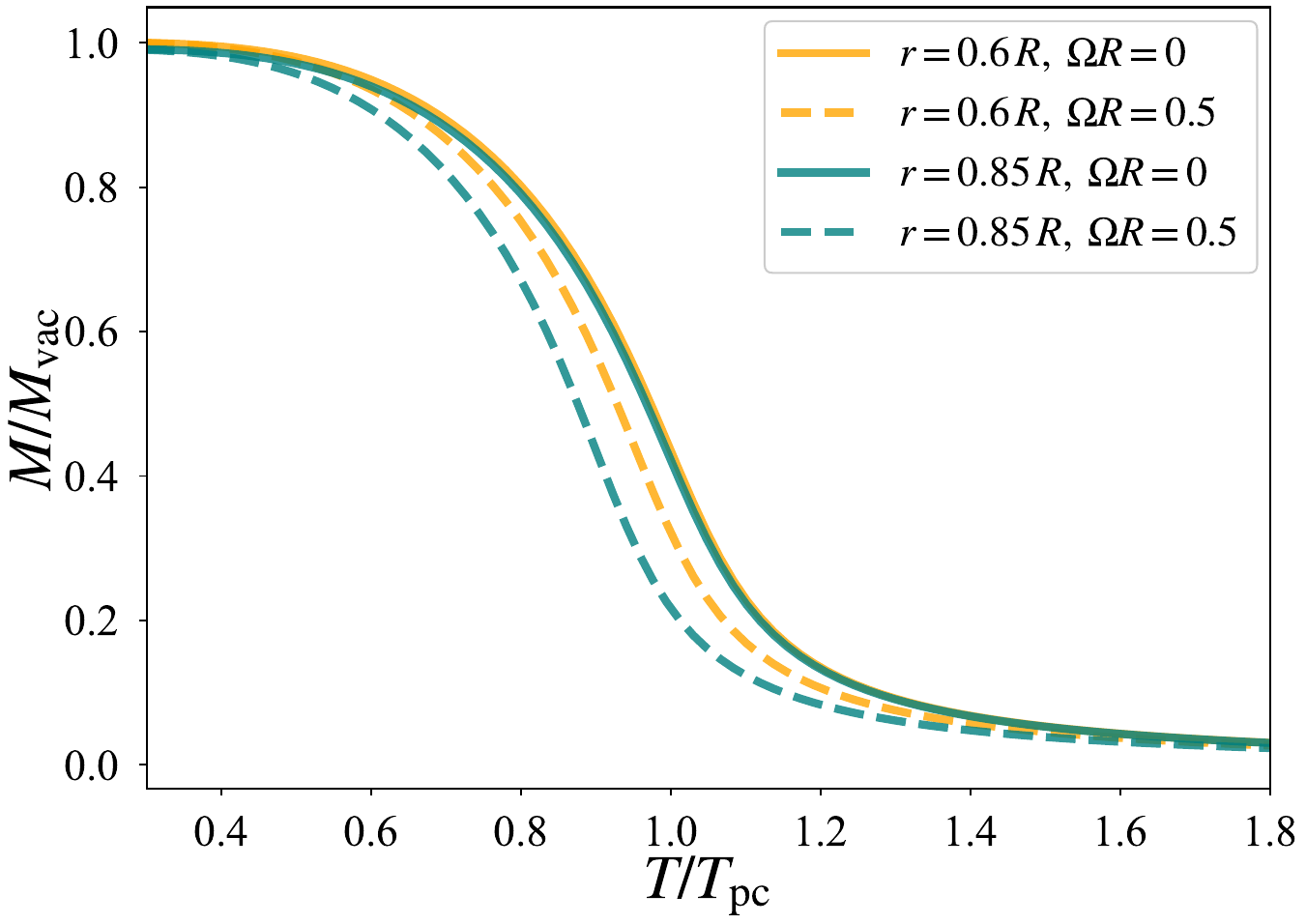}
\end{center}
\caption{
The temperature dependence of the dynamical quark mass at fixed $r$.
The temperature is normalized by the pseudocritical temperature evaluated in the absence of rotation and boundary effects.
}
\label{Tem_dynamical_mass_rotate}
\end{figure}

In addition to the $r$-dependence, we also present the temperature dependence of the dynamical quark mass at fixed $r$ in Fig.~\ref{Tem_dynamical_mass_rotate}, in order to clearly illustrate the rotational effect on the thermal chiral phase transition.
This figure shows that the rotational effect enhances the suppression of the dynamical quark mass at finite temperature, indicating that the rotational effect promotes the restoration of chiral symmetry and becomes more pronounced near the boundary.

Based on the above evaluations, we proceed to numerically compute the susceptibility functions in the following.

%%%%%%%%%%%%%%%%%%%%%%%%%%%%%%%%%%%%%%%%%%%%
\subsection{Meson and topological susceptibilities}
\begin{figure}
\begin{tabular}{cc}
\begin{minipage}{0.5\hsize}
\begin{center}
    \includegraphics[width=1\columnwidth]{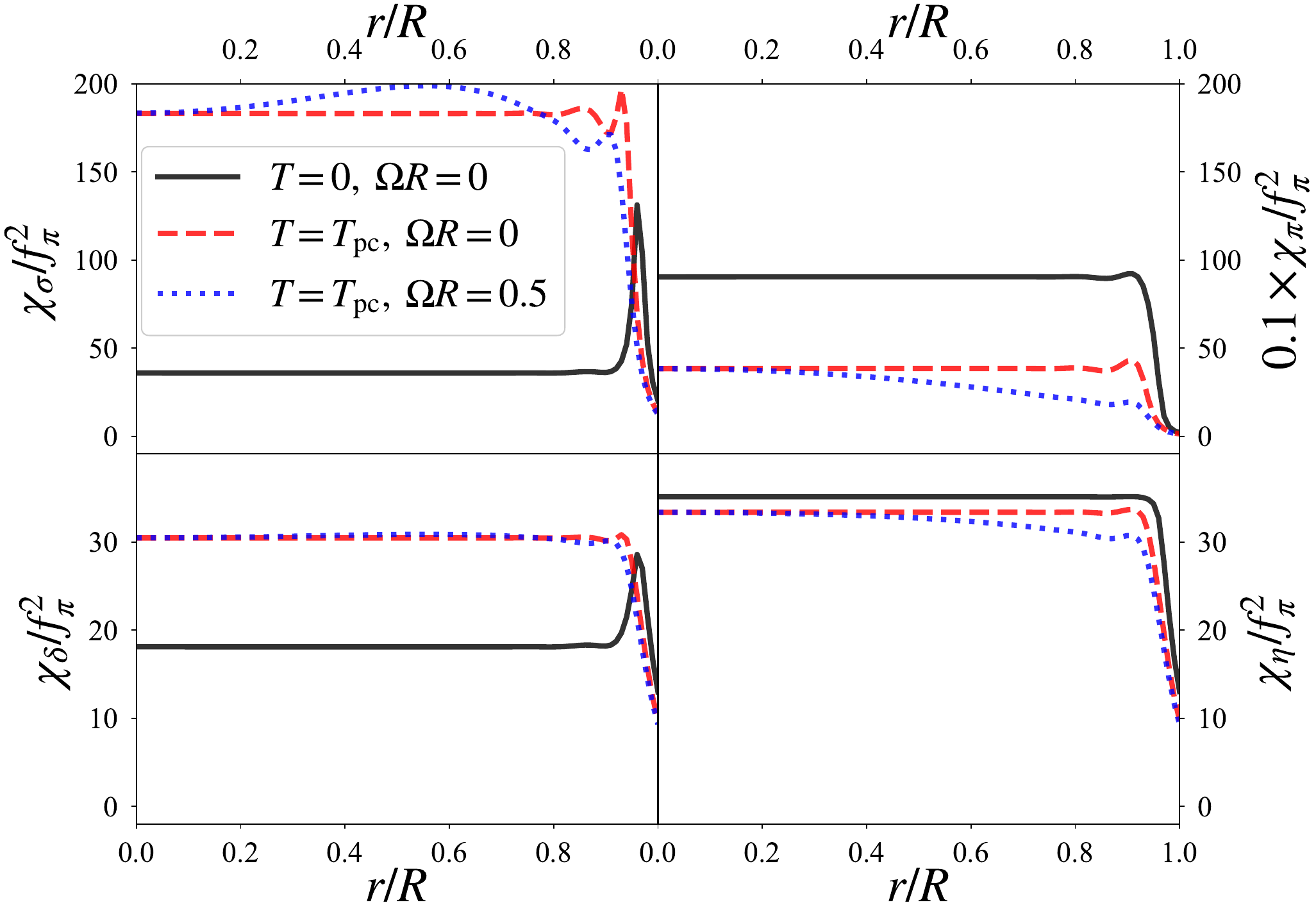}
    \subfigure{(a)}
\end{center}
\end{minipage}
\begin{minipage}{0.5\hsize}
\begin{center}
    \includegraphics[width=1\columnwidth]{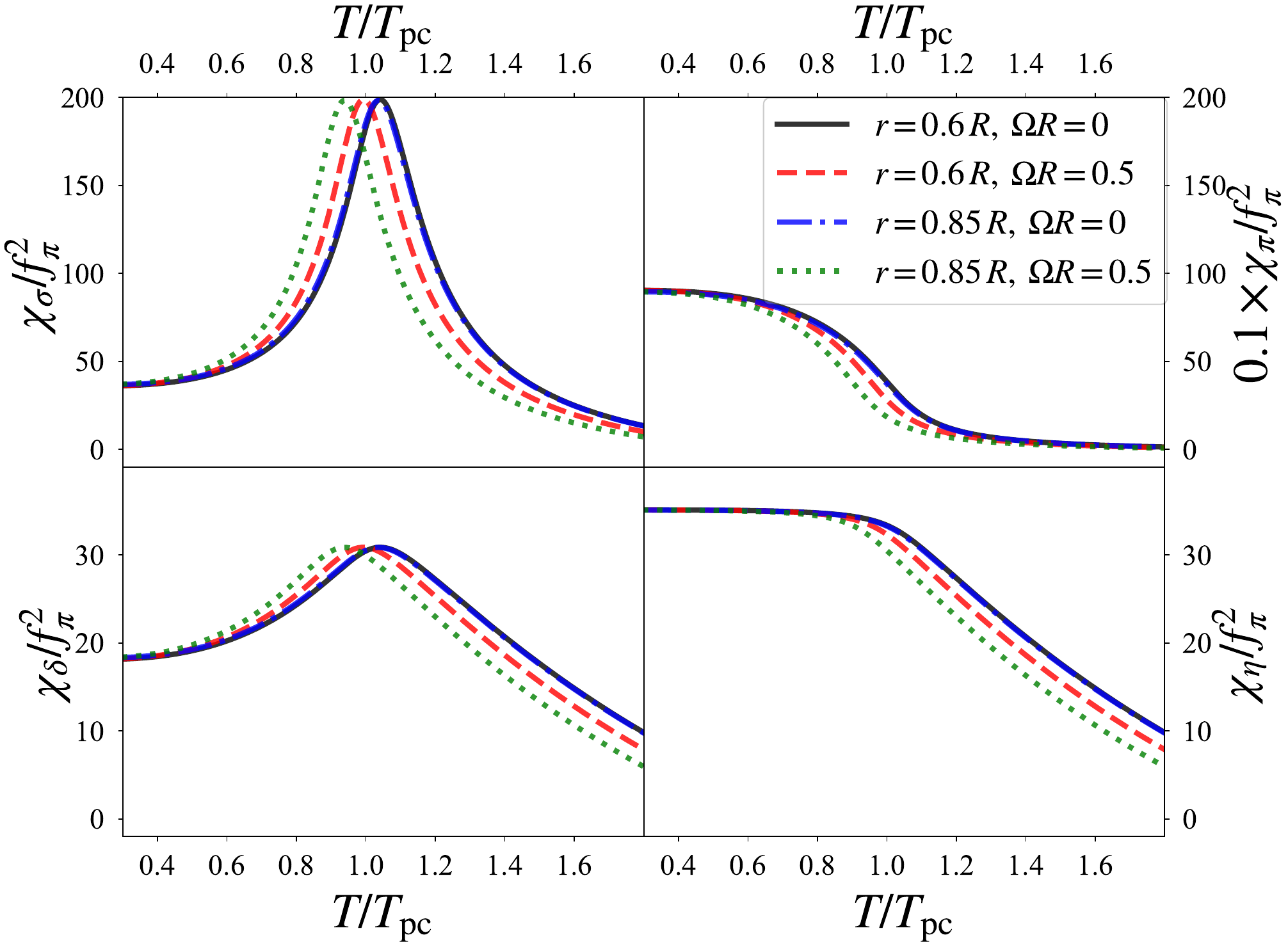}
    \subfigure{(b)}
\end{center}
\end{minipage}
\end{tabular}
\caption{
The rotational effect on the meson susceptibilities normalized by the square of the pion decay constant ($f_\pi=92.4\,{\rm MeV}$). 
(a): the $r$-dependence of meson susceptibilities at fixed temperature and
(b): the temperature dependence of meson susceptibilities at fixed $r$.
}
\label{chi_meson_r_T}
\end{figure}

In Fig.~\ref{chi_meson_r_T}, we numerically evaluate the meson susceptibilities under rotation.
Panel~(a) illustrates its $r$-dependence at fixed temperature, both with and without the rotational effect, clearly showing the difference in the $r$-dependence between the scalar meson susceptibilities $(\chi_{\sigma,\delta})$  and pseudoscalar meson susceptibilities $(\chi_{\pi,\eta})$.
This difference comes from the analytical one-loop results evaluated in Eq.~\eqref{one_loop_analytical}. 
Looking at the details of each meson susceptibilities,
the sigma meson susceptibility, which corresponds to the isosinglet scalar channel, remains approximately constant in the relevant region ($r<0.9\,R$) at zero temperature. 
At finite temperature, its magnitude becomes larger, but it still exhibits constant behavior. When including the rotational effect, the sigma meson susceptibility increases with $r$ and then decreases near the boundary in order to satisfy the boundary condition.
Similarly, the delta meson susceptibility, which corresponds to the isotriplet scalar channel, shows a qualitatively similar behavior to that of the isosinglet channel $\chi_\sigma$.
However, compared to the sigma meson, the delta meson susceptibility has a smaller overall magnitude and is less affected by the rotational effect. Since the meson susceptibilities are roughly inversely proportional to the square of the related meson masses (see, i.e., the linear sigma model result~\cite{Kawaguchi:2023olk}), the overall magnitude of the delta meson susceptibility, associated with a heavier scalar meson, is suppressed, and the effect of the angular velocity diminishes for the heavier scalar meson.

The pseudoscalar meson susceptibilities also show comparable behavior between isotriplet and isosinglet pseudoscalar mesons.
In contrast to the scalar meson susceptibilities, the overall magnitude of the pion susceptibility is suppressed with increasing the temperature, and decreases with increasing $r$ in the presence of rotation. This behavior aligns with that of the dynamical quark mass. Indeed, the pion susceptibility is directly connected to the quark condensate via the WT identity associated with the $SU(2)_A$ symmetry, as shown in Eq.~\eqref{WTISU2}. 
For the isosinglet channel, the eta meson susceptibility exhibits a trend similar to that of the isotriplet channel $\chi_\pi$, but with an overall smaller magnitude and a reduced sensitivity to the rotational effect. This is again attributed to the larger mass of the eta meson compared to the pion mass.

We also depict the temperature dependence of the meson susceptibilities at fixed $r$ in panel~(b) of Fig.~\ref{chi_meson_r_T}. One can see that the rotational effect acts as a catalyzer, enhancing the thermal suppression of the meson susceptibilities, and its influence becomes more pronounced near the boundary.

\begin{figure} %htbp
\begin{tabular}{cc}
\begin{minipage}{0.5\hsize}
\begin{center}
    \includegraphics[width=0.95\columnwidth]{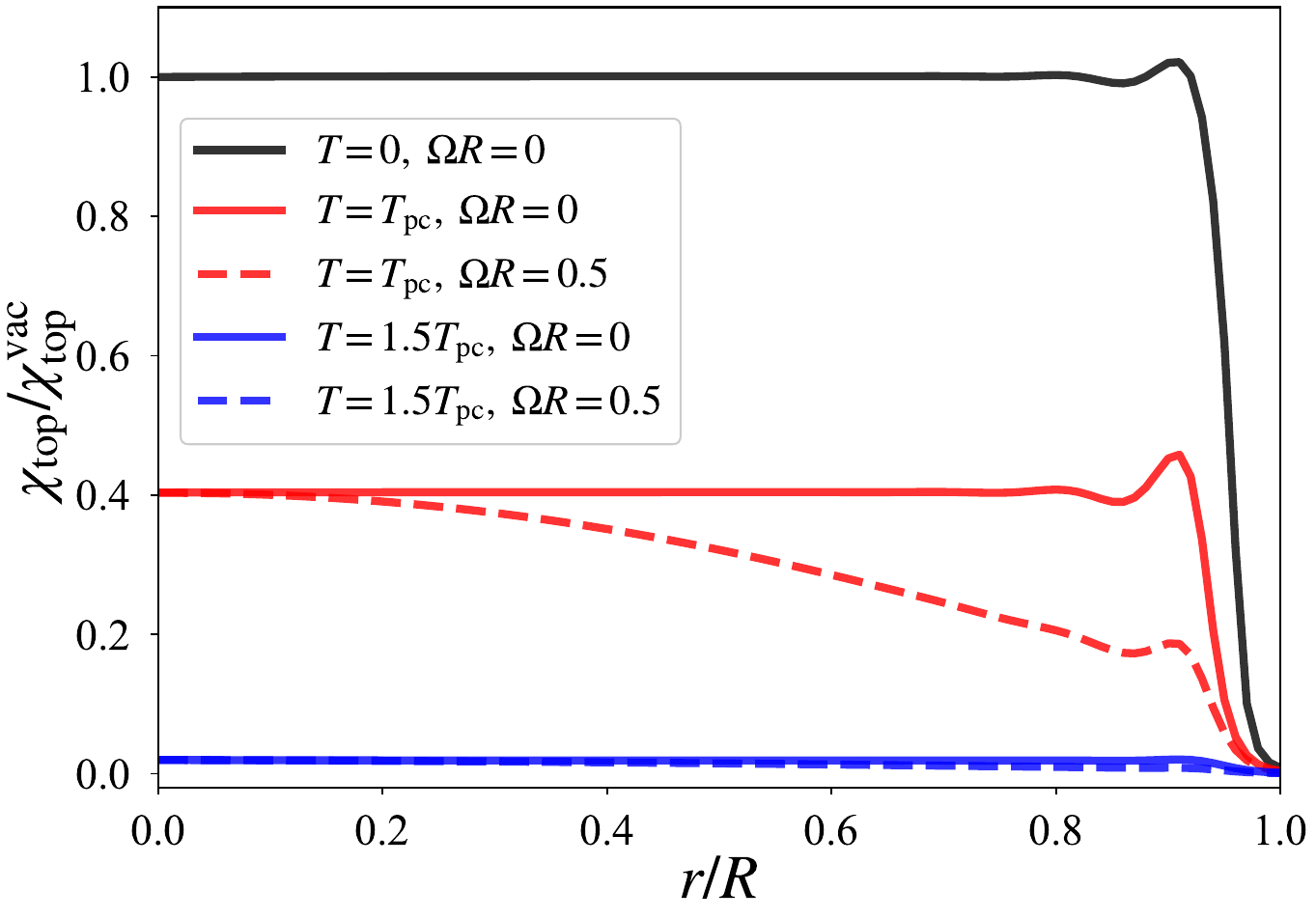}
    \subfigure{(a)}
\end{center}
\end{minipage}
\begin{minipage}{0.5\hsize}
\begin{center}
    \includegraphics[width=0.95\columnwidth]{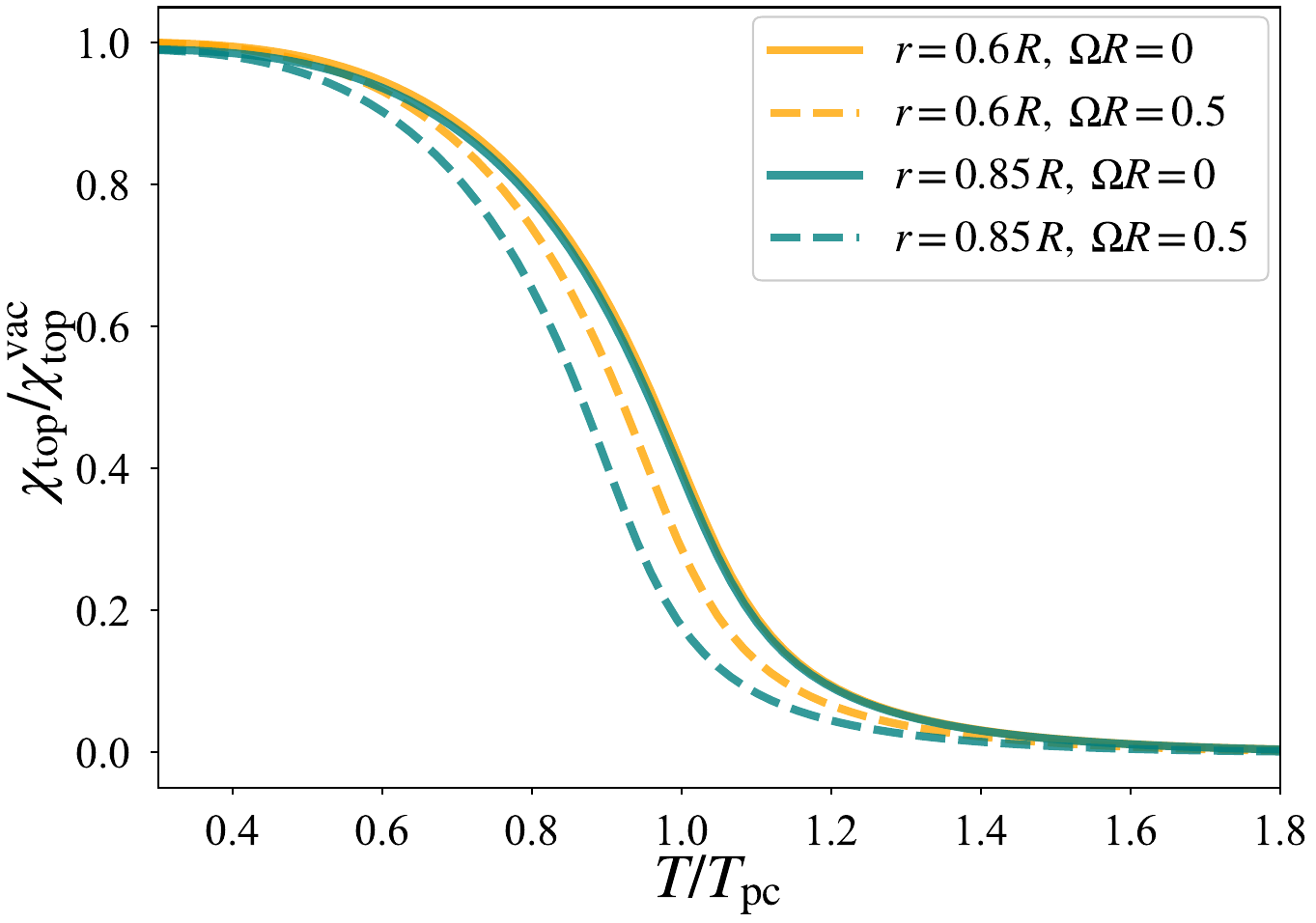}
    \subfigure{(b)}
\end{center}
\end{minipage}
\end{tabular}
\caption{
The rotational effect on the topological susceptibility normalized by its vacuum value $\chi^{\rm vac}_{\rm top}$. 
(a): the $r$-dependence of $\chi_{\rm top}$ and
(b): the temperature dependence of $\chi_{\rm top}$.
}
\label{chi_top_r_T}
\end{figure}

Besides, we evaluate the topological susceptibility under rotation by the analytical expression in Eq.~\eqref{chi_top_alt} with the numerical results of the pseudoscalar meson susceptibilities obtained above.
Fig.~\ref{chi_top_r_T} presents the $r$-dependence in panel~(a) and the temperature dependence in panel~(b).
These dependences closely resemble the behaviors of the pseudoscalar meson susceptibilities shown in Fig.~\ref{chi_meson_r_T}.
From these results, the role of the rotational effect on the topological susceptibility can be understood as follows:
it further facilitates the thermal suppression of the topological susceptibility and becomes more pronounced near the boundary, following a trend similar to that of the pseudoscalar meson susceptibilities.

In addition, the behavior of the topological susceptibility in a rotating system can also be understood as follows.
As shown in Fig.~\ref{chi_meson_r_T}, the $\eta$ meson susceptibility is relatively small, whereas the pion susceptibility is significantly larger, leading to $\chi_\pi$ being the dominant contribution to the topological susceptibility. Notably, the pion susceptibility can be expressed in terms of the quark condensate via the WT identity in Eq.~\eqref{WTISU2}. 
Therefore, the behavior of the topological susceptibility reflects that of the quark condensate under rotation.
Indeed, a similar correspondence between the behavior of the topological susceptibility and that of the quark condensate has also been observed in other contexts, such as finite temperature systems~\cite{Ruivo:2011fg,Jiang:2012wm,GomezNicola:2019myi,Kawaguchi:2020kdl,Kawaguchi:2020qvg,Cui:2021bqf,Cui:2022vsr,Fejos:2025oxi} (for details of functional renormalization group analysis, see also Ref.~\cite{Fejos:2016hbp}) and finite-density systems~\cite{Li:2022dry,Kawaguchi:2023olk}.

%%%%%%%%%%%%%%%%%%%%%%%%%%%%%%%
\subsection{Baryon number susceptibility}
\begin{figure} 
\begin{tabular}{cc}
\begin{minipage}{0.5\hsize}
\begin{center}
    \includegraphics[width=0.95\columnwidth]{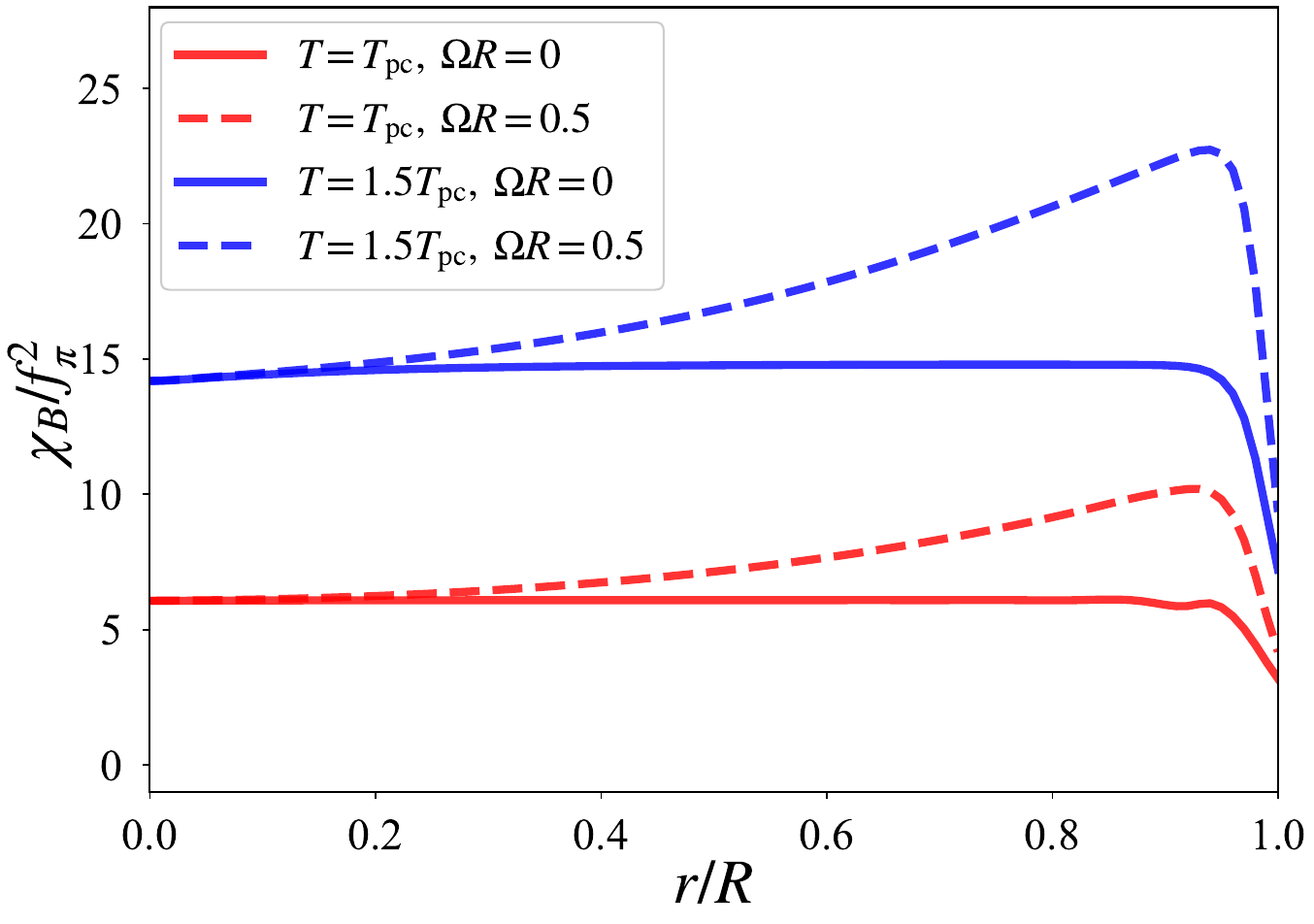}
    \subfigure{(a)}
\end{center}
\end{minipage}
\begin{minipage}{0.5\hsize}
\begin{center}
    \includegraphics[width=0.95\columnwidth]{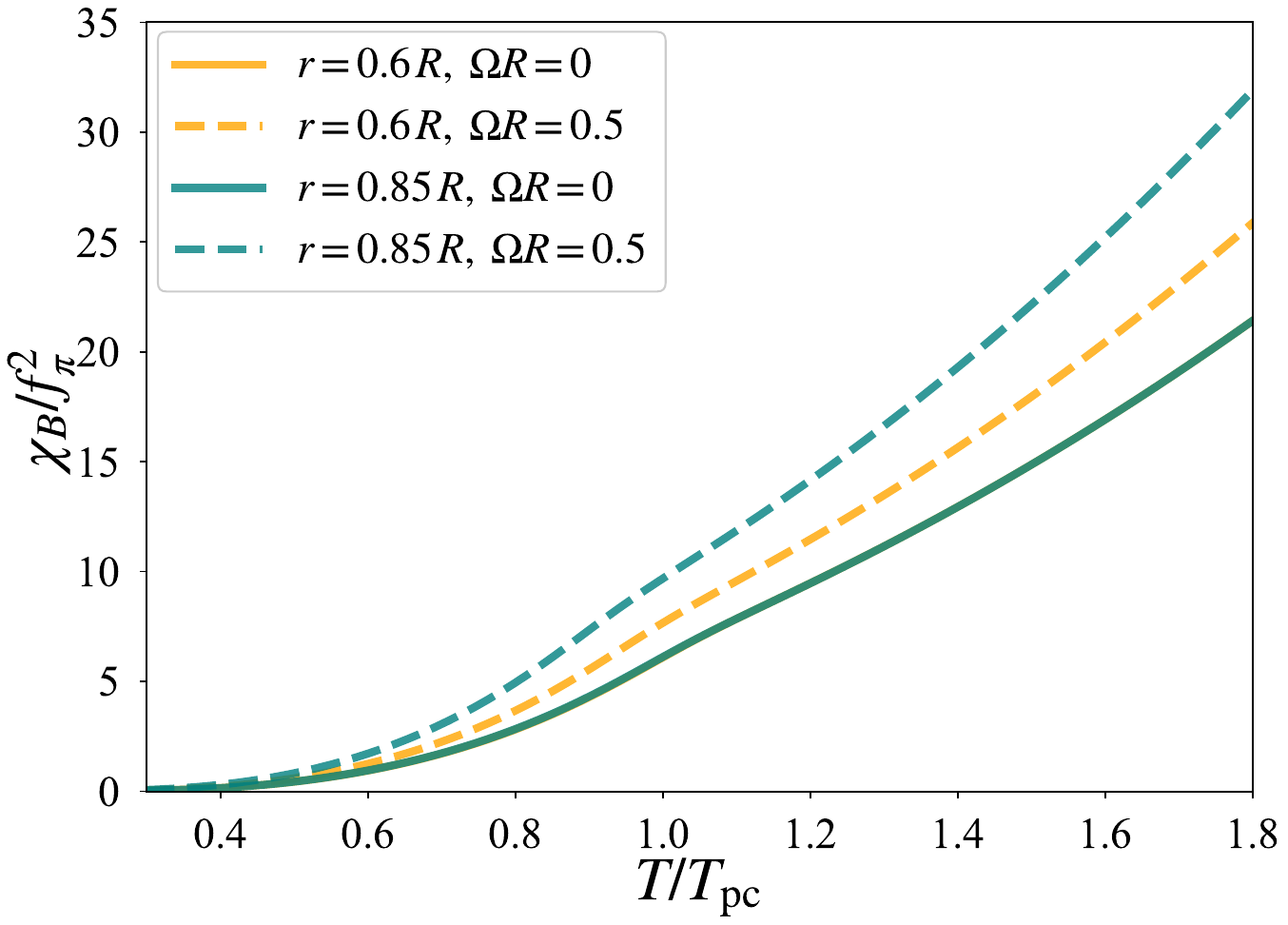}
    \subfigure{(b)}
\end{center}
\end{minipage}
\end{tabular}
\caption{
The rotational effect on the baryon number susceptibility normalized by the square of the pion decay constant. 
(a): the $r$-dependence of $\chi_{\rm B}$ and
(b): the temperature dependence of $\chi_{\rm B}$.
}
\label{bary_sus_numerical}
\end{figure}

In this subsection, we exhibit the baryon number susceptibility in Fig~\ref{bary_sus_numerical}. 
The baryon number susceptibility can take finite values at finite temperatures and it remains constant within the relevant region of $r \leq 0.9\,R$ in the absence of the rotational effect, as shown in panel~(a).  When including the rotational effect, the baryon number susceptibility gets the $r$ dependence, which is suppressed as it approaches the boundary.
This suppression, driven by the rotational effect at finite temperature, is also clearly visible in panel~(b). 

\begin{figure}
\begin{tabular}{cc}
\begin{minipage}{0.5\hsize}
\begin{center}
\includegraphics[width=0.95\columnwidth]{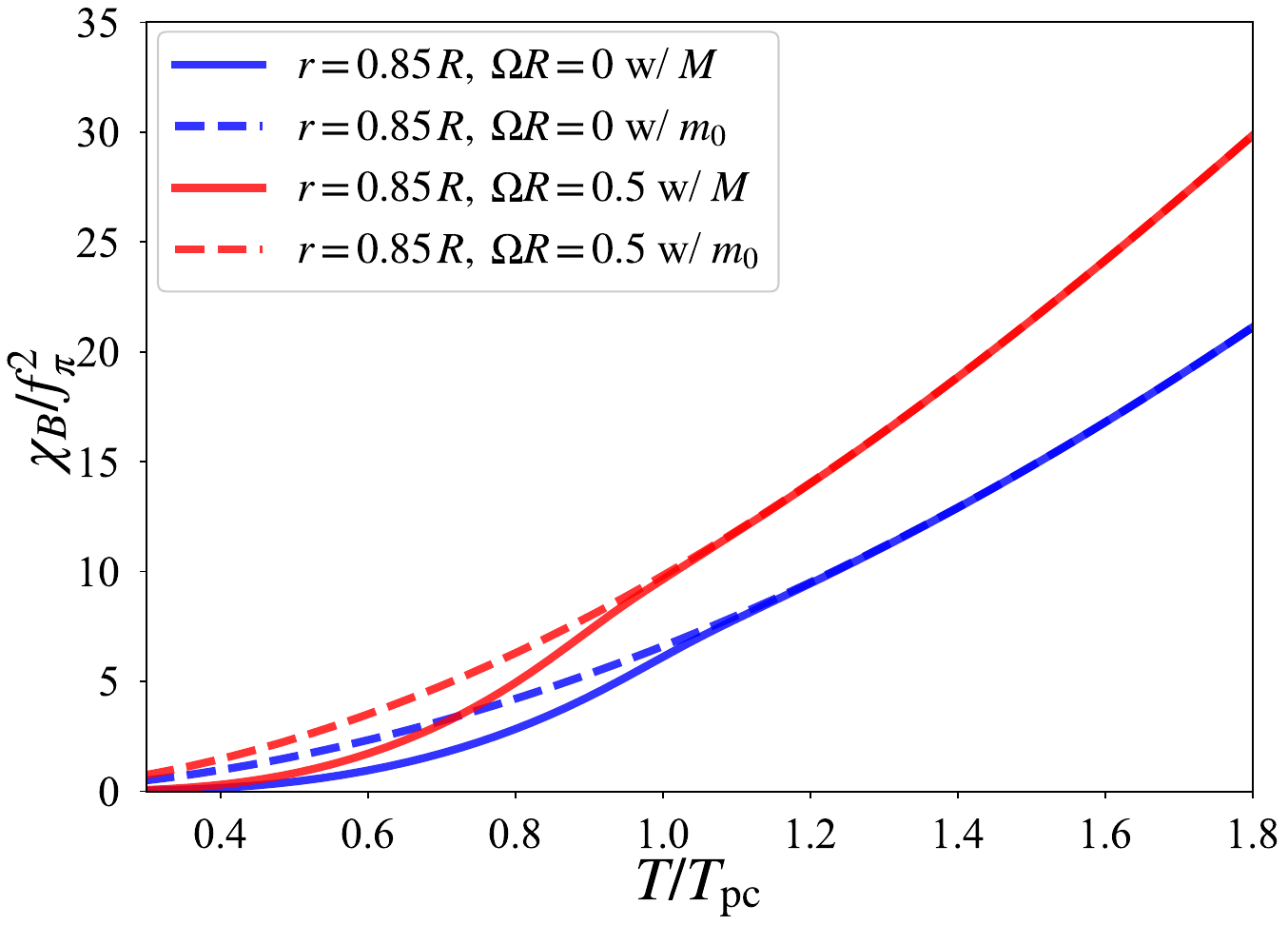}
    \subfigure{(a)}
\end{center}
\end{minipage}
\begin{minipage}{0.5\hsize}
\begin{center}
    \includegraphics[width=0.95\columnwidth]{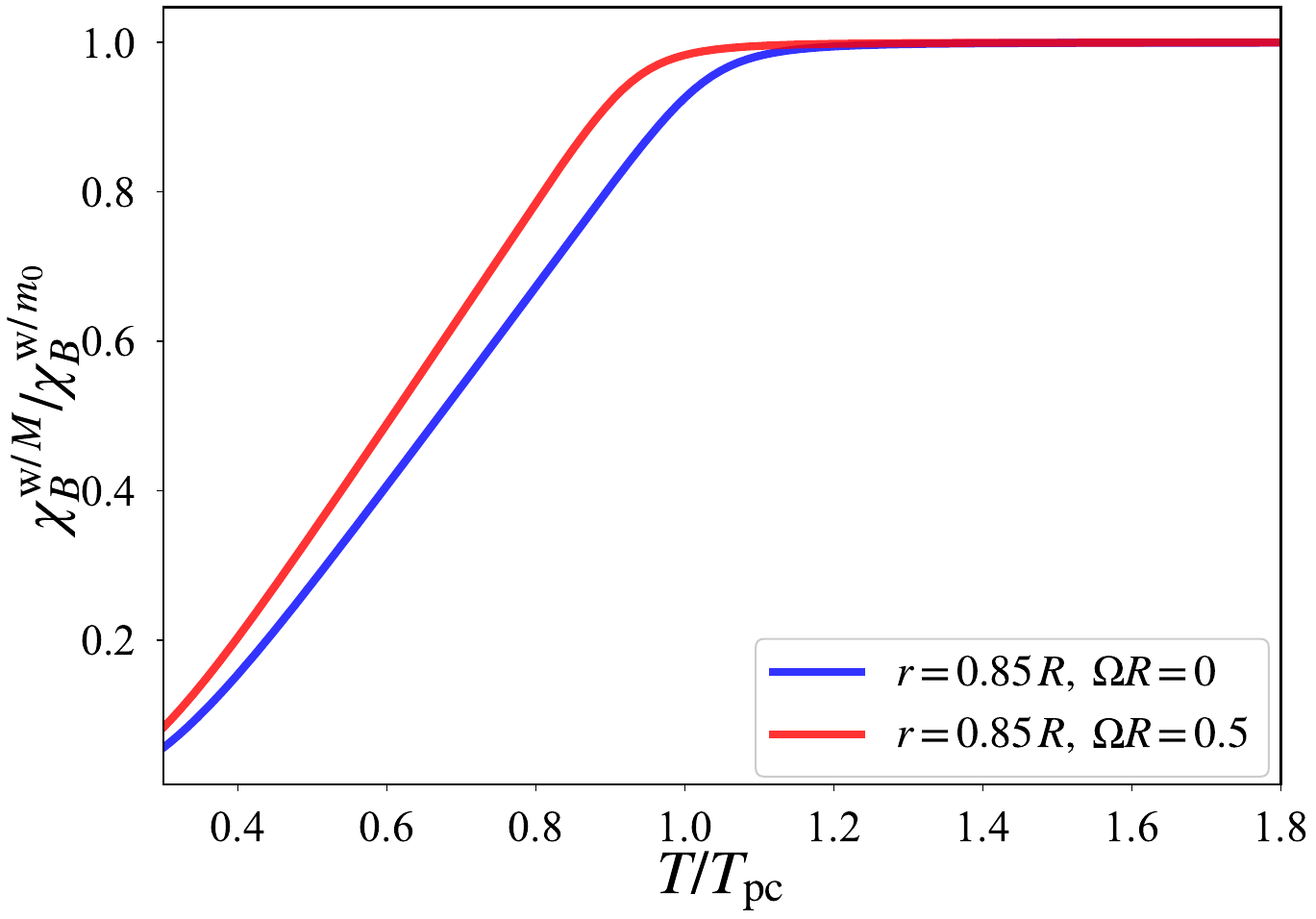}
    \subfigure{(b)}
\end{center}
\end{minipage}
\end{tabular}
\caption{
(a) Comparison between the baryon number susceptibilities in the interacting cases (solid lines, $\chi_B^{{\rm w}/M}$) and noninteracting cases (dashed lines, $\chi_B^{{\rm w}/m_0}$). 
(b) Baryon number susceptibility normalized by that in the noninteracting case.
}
\label{comp_m0}
\end{figure}

In Fig.~\ref{comp_m0}, we numerically compute the baryon number susceptibility.
To investigate the influence of the chiral phase transition, we compare the susceptibilities in the interacting and noninteracting cases.
The former (latter) is denoted by $\chi_B^{{\rm w}/M}$ ($\chi_B^{{\rm w}/m_0}$) and plotted with the solid (dashed) lines;
the latter is evaluated by replacing the dynamical quark mass $M$ in the Fermi distribution function $n_F(E)$ of Eq.~\eqref{baryon_sus} with the current quark mass $m_0$.
In the absence of the rotational effect, 
while for $T\lesssim T_{\rm pc}$, there are discrepancies between the two cases, as depicted in panel~(a), such discrepancies vanish for $T\gtrsim T_{\rm pc}$.
This is because the dynamical quark mass approaches the current quark mass at high temperatures.
To clearly capture this trend, in panel~(b) we plot $\chi_B^{{\rm w}/M}/\chi_B^{{\rm w}/m_0}$, which is the susceptibility normalized by that in the noninteracting case.
When including the rotational effects, this convergence shifts to lower-temperature regions, as shown in panel~(b). The baryon number susceptibility is correlated with the chiral restoration affected by the rotational effect.

%%%%%%%%%%%%%%%%%%%%%%%%%%%%%%%
\subsection{Moment of inertia}
\begin{figure} 
\begin{tabular}{cc}
\begin{minipage}{0.5\hsize}
\begin{center}
    \includegraphics[width=0.95\columnwidth]{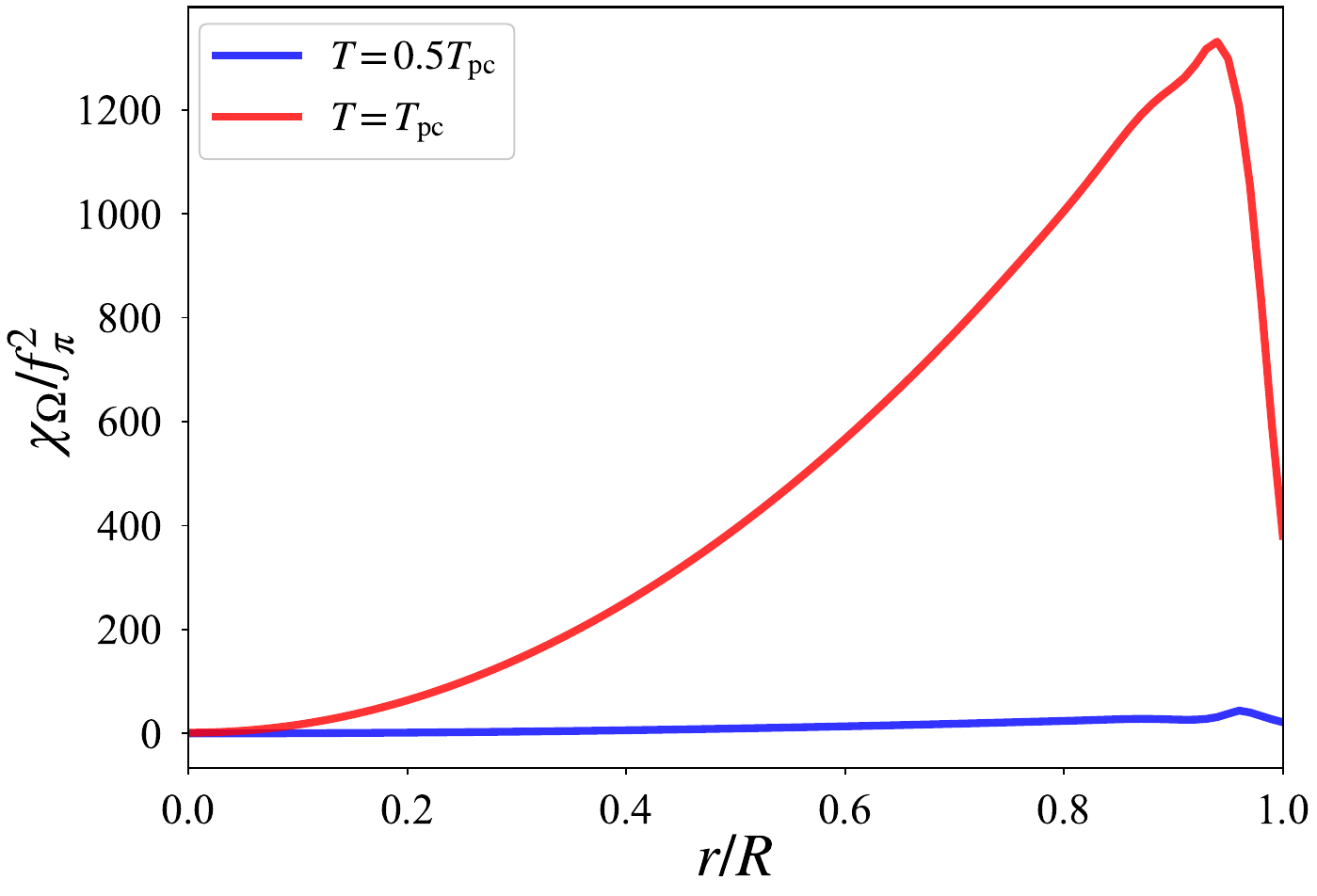}
    \subfigure{(a)}
\end{center}
\end{minipage}
\begin{minipage}{0.5\hsize}
\begin{center}
    \includegraphics[width=0.95\columnwidth]{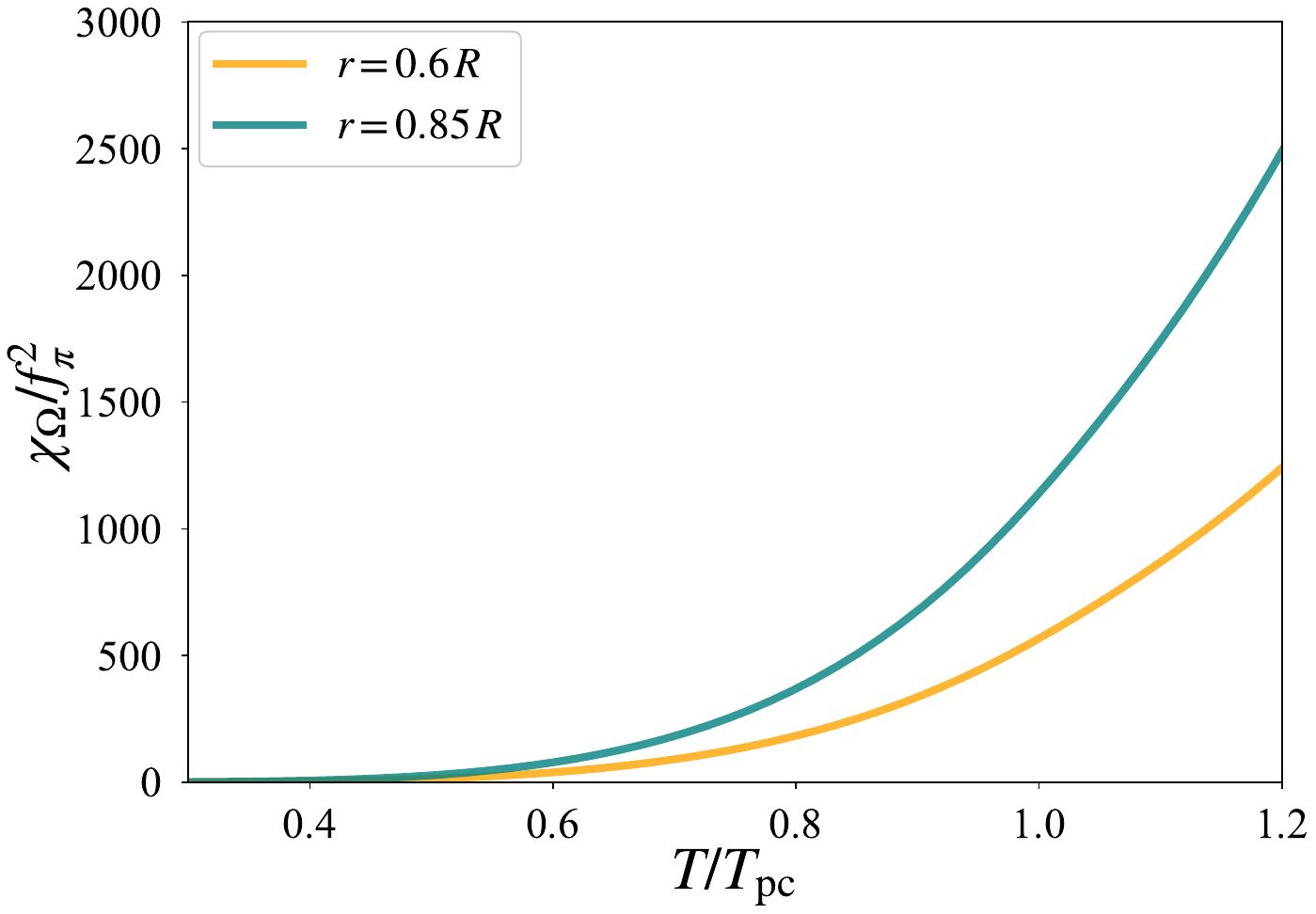}
    \subfigure{(b)}
\end{center}
\end{minipage}
\end{tabular}
\caption{
(a) The $r$-dependence of the moment of inertia normalized by the square of the pion decay constant.
(b) The temperature dependence of the normalized moment of inertia at fixed $r$. 
}
\label{moment_in_nume}
\end{figure}

In Fig.~\ref{moment_in_nume}, we show numerical results of the moment of inertia~\eqref{def_MoI}, which is independent of $\Omega$.
Panel~(a) is the radial-coordinate dependence, and the blue and red lines correspond to the symmetry-broken and symmetry-restored cases, respectively.
First, we focus on the latter. While the moment of inertia is tiny around $r=0$, it grows for larger $r$ and reaches a peak around $r=0.9\,R$.
This behavior stems from the contributions from the modes that has a large total angular momentum $j=l+1/2$, and thus spatially locate around a large $r$.
Such a different location of each mode is the reason why the moment of inertia exhibits the clear $r$-dependence even at $\Omega=0$, while other susceptibility functions are insensitive to $r$ at $\Omega=0$.
On the other hand, the blue line in panel~(a) exhibits the small value irrelevantly to $r$, because the fermionic distribution is suppressed by a huge dynamical mass.
This observation is supported by panel~(b), where the moment of inertia is a monotonically increasing function of $T$.
We note that the difference between the green and yellow lines again stems from the contribution from large $j$-modes.

\begin{figure} 
\begin{center}
    \includegraphics[width=0.47\columnwidth]{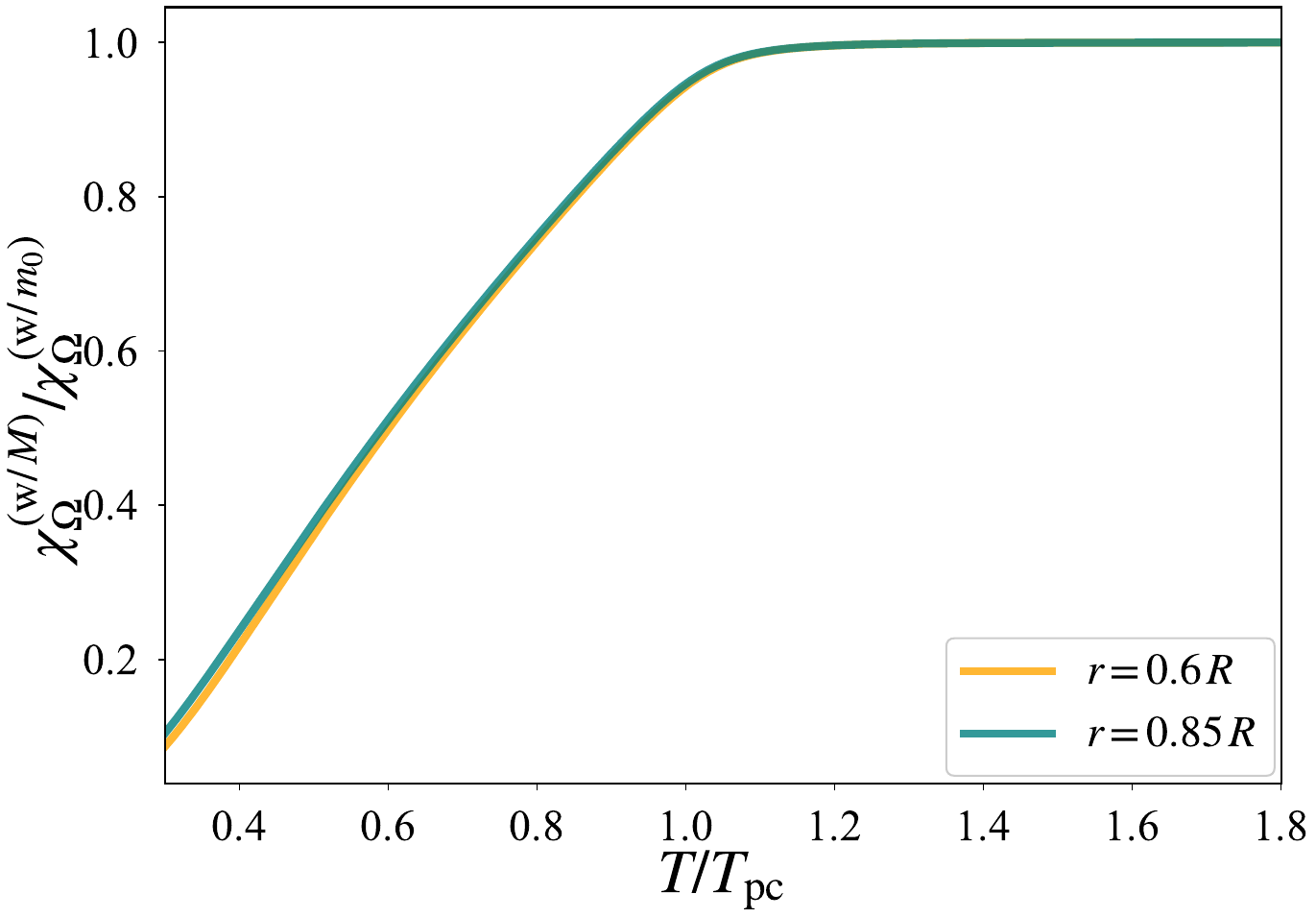}
    % \subfigure{(a)}
\end{center}
\caption{
Moment of inertia normalized by that in the noninteracting case.
}
\label{moment_in_nume_ratio}
\end{figure}

Similarly to the baryon number susceptibility, in Fig.~\ref{moment_in_nume_ratio} we show the moment of inertia normalize by that in the noninteracting case, $\chi_\Omega^{{\rm w}/M}/\chi_\Omega^{{\rm w}/m_0}$.
The convergent behavior to unity at $T\gtrsim T_{\rm pc}$ implies that the moment of inertia has the sensitivity to the chiral phase transition, similarly to the baryon number susceptibility.
Hence, the moment of inertia is one of the indicators of the transition.
We note that this normalized moment of inertia is almost independent of $r$, for the following reason.
While the $r$-dependence of $\chi_\Omega^{{\rm w}/M}$ comes from the dynamical mass $M(r)$ and the Bessel function $J_l(p_{l,k}r)$, that of $\chi_\Omega^{{\rm w}/m_0}$ comes only from $J_l(p_{l,k}r)$.
As shown in Fig.~\ref{dynamical_mass_rotate}, however, $M(r)$ at $\Omega=0$ is almost independent of $r$.
As a result, the $r$-dependence of $\chi_\Omega^{{\rm w}/M}$ is the same as that of $\chi_\Omega^{{\rm w}/m_0}$, leading to the $r$-independent ratio $\chi_\Omega^{{\rm w}/M}/\chi_\Omega^{{\rm w}/m_0}$.

%%%%%%%%%%%%%%%%%%%%%%%%%%%%%%%%%%%%%%%%%5
\section{Discussion and summary}
\label{summary}

In this study, we investigated the
fermionic two-point correlation functions in the rotating finite-size cylinder, focusing particularly on the susceptibility functions. Taking into account the violation of translational symmetry under rotation, we began by formulating the Dirac propagator using the Fourier-Bessel basis. 
To specifically investigate the rotational effect on susceptibility functions in an interacting theory, we employed the two-flavor NJL model.
Within the mean-field approximation together with the local density approximation, the quark propagator in the rotating frame takes the same form as the usual free one, but involving the radial-coordinate dependent dynamical quark mass.
Based on this formulation, we also analyze the meson, topological and baryon number susceptibilities as well as the moment of inertia.
Under rotation, the violation of the translational symmetry seems to be incompatible to the local density approximation, and yields a complication in the computation of meson susceptibilities.
However, we have found that thanks to a WT identity, the meson susceptibilities can still be expressed in a form analogous to that evaluated by the conventional ring-diagram resummation.

We also numerically evaluated the susceptibility functions.
The susceptibilities associated with the lightest scalar and pseudoscalar mesons are more sensitive to rotational effects than those corresponding to heavier meson channels.
Rotation amplifies the thermal suppression of these susceptibilities, with the effect becoming particularly pronounced near the boundary of the system.
The topological susceptibility, evaluated through pseudoscalar meson susceptibilities, exhibits similar behavior and signals the effective restoration of the $U(1)_A$ symmetry at finite temperature, especially near the boundary of the rotating cylinder.
We have also evaluated the baryon number susceptibility, which increases with radial distance under rotation and reflects the chiral restoration near the pseudocritical temperature.
Although the moment of inertia shows a peak near the boundary even without rotation, it also serves as an indicator of the chiral phase transition, similar to the baryon number susceptibility.

As a side remark, we note that a recent study has computed susceptibilities in the context of investigating the critical end point in the QCD phase diagram~\cite{Chen:2024hki}.
The authors evaluate the quark–antiquark spin correlation, which is formulated in terms of quantities analogous to the baryon number susceptibility and the moment of inertia.
In their approach, separate chemical potentials and angular velocities are introduced for quarks and antiquarks, from which they derive corresponding susceptibility functions.
As a result of this independent treatment of quarks and antiquarks, the physical quantities they obtain differ from the baryon number susceptibility and moment of inertia considered in our analysis.
Besides, the susceptibility functions evaluated in this paper is the local one, unlike that in Ref.~\cite{Chen:2024hki}.

Although we have adopted the mean field approximation supplemented by the local density approximation, this study presents the first analysis of the susceptibility functions of rotating quark fields, properly taking into account the violation of the translational invariance.
The susceptibility functions are measurable in lattice QCD simulations even under finite angular velocity, employing techniques such as the imaginary angular velocity
method~\cite{Braguta:2022str} and Taylor expansion analysis~\cite{Yang:2023vsw}.
Furthermore, the moment of inertia evaluated at zero angular velocity is not only a computable quantity without suffering from a sign problem, but also an essential ingredient characterizing the rotational response on the vacuum~\cite{Braguta:2023yjn,Mameda:2023sst}.
An important remark is that moment of inertia is strongly affected by the geometry of systems, as well as boundary conditions.
For instance, the orbital contribution is always washed out in the Cartesian coordinate system with the spatially periodic boundary condition, unlike our analysis in this paper.
Including the treatment of such an inhomogeneity in lattice gauge theory, more comprehensive discussions could bridge a gap between lattice simulations of rotating QCD matter and other approaches.

%%%%%%%%%%%%%%%%%%%%%%%%%%%%%%
%\begin{acknowledgments}
\section*{Acknowledgment}
This work is supported by RFIS-NSFC under Grant No.~W2433019, and JSPS KAKENHI Grant No.~24K17052. 
%\end{acknowledgments}

\bibliographystyle{JHEP}
\bibliography{reference}

\end{document}